\documentclass[aps,pre,footinbib,showpacs,twocolumn]{revtex4-1}

\usepackage{amssymb,latexsym,mathrsfs}
\usepackage{amsfonts}
\usepackage{graphicx}
\usepackage{lmodern}
\usepackage{fix-cm}
\usepackage{xfrac}
\usepackage{color}    
\usepackage{textcomp}
\usepackage{amsmath}
\usepackage{subfigure}
\usepackage{float}
\usepackage{datetime}
\usepackage{hyperref}
\hypersetup{
    colorlinks=true,
    linkcolor=red,
    filecolor=magenta,      
    urlcolor=cyan,
}
\usepackage{tabularx,ragged2e,booktabs}

\newcommand{\be}{\begin{equation}}
\newcommand{\ee}{\end{equation}}
\newcommand{\bearr}{\begin{array}}
\newcommand{\enarr}{\end{array}}

\def\bea{\begin{eqnarray}}
\def\eea{\end{eqnarray}}
\def\ba{\begin{array}}
\def\ea{\end{array}}

\definecolor{dgreen}{rgb}{0,0.7,0}

\begin{document}

\title{Dynamical regimes of finite temperature discrete nonlinear Schr\"{o}dinger chain}

\author{Amit Kumar Chatterjee}\email{amit.chatterjee@icts.res.in}
\affiliation{International Centre for Theoretical Sciences, Tata Institute of Fundamental Research, Bengaluru -- 560089, India}

\author{Manas Kulkarni}\email{manas.kulkarni@icts.res.in}
\affiliation{International Centre for Theoretical Sciences, Tata Institute of Fundamental Research, Bengaluru -- 560089, India}

\author{Anupam Kundu}\email{anupam.kundu@icts.res.in}
\affiliation{International Centre for Theoretical Sciences, Tata Institute of Fundamental Research, Bengaluru -- 560089, India}


\begin{abstract}
We show that the one dimensional discrete nonlinear Schr\"{o}dinger chain (DNLS) at finite temperature has three different dynamical regimes (ultra-low, low and high temperature regimes). This has been established via (i) one point macroscopic thermodynamic observables (temperature $T$, energy density $\epsilon$ and the relationship between them), (ii) emergence and disappearance of an additional almost conserved quantity (total phase difference) and (iii) classical out-of-time-ordered correlators (OTOC) and related quantities (butterfly speed and Lyapunov exponents). The crossover temperatures $T_{l-ul}$ (between low and ultra-low temperature regimes) and $T_{h-l}$ (between high and low temperature regimes) extracted from these three different approaches are consistent with each other.  
The analysis presented here is an important step forward towards the understanding of DNLS which is ubiquitous in many fields and has a non-separable Hamiltonian form. Our work also shows that the different methods used here can serve as important tools to identify dynamical regimes in other interacting many body systems.
\end{abstract}

\maketitle


\section{Introduction}
\label{intro}
The discrete nonlinear Schr\"{o}dinger chain (DNLS) exhibits a plethora of interesting mathematical and physical features, and has a wide range of applicability in real systems \cite{Ablowitz_2004, Kevrekidis_2001,Hennig_1999}. A key feature of this model is its experimental realizability. For example, the solitary waves found mathematically from the DNLS equations \cite{Christodoulides_1988} have been observed experimentally in nonlinear optical waveguide arrays \cite{Eisenberg_1998,Eisenberg_2000,Morandotti_1999}. The importance of DNLS has also been revealed in various fields, ranging from transport in biological systems \cite{Davydov_1973, Davydov_1981} to condensed matter systems like localized modes in anharmonic crystals \cite{Sievers_1988}, soliton formation in semiconducting polymers \cite{Su_1979}, Bose-Einstein condensates \cite{Trombettoni_2001} to name a few. 

From the perspective of statistical mechanics, the {\it non-separable} structure of the DNLS Hamiltonian (i.e. the Hamiltonian is not in the form of a sum of kinetic energy and potential energy) being non-trivial, naturally demands for extensive studies regarding the thermalization of the system. In this connection, it has been elaborately discussed in Ref.~\onlinecite{Rasmussen_2000} that a Gibbs measure is applicable for the one-dimensional DNLS  chain at positive temperatures. However, it is possible to prepare the DNLS at negative temperatures \cite{Iubini_2013_NJP}, where one can observe 
localized breather-like modes \cite{Rasmussen_2000, Iubini_2013_NJP,Iubini_2014} which cannot be described by Gibbs measure. 

The non-equilibrium steady states of the DNLS model has been investigated by adapting suitable Monte Carlo \cite{Iubini_2012} and Langevin thermostats \cite{Iubini_2013}. This non-integrable model has two conserved quantities (norm and energy) and it exhibits rich coupled transport phenomena along with interesting non-monotonous energy and density profiles \cite{Iubini_2012}. 

Recently, there has been an interesting observation concerning the DNLS in equilibrium. It has been revealed that the one-dimensional DNLS exhibits three different dynamical regimes \cite{Mendl_2015}, namely the {\it high temperature, low temperature and ultra-low temperature regimes}. Notably, the observables used to differentiate the three dynamical regimes in Ref.~\onlinecite{Mendl_2015}, are the {\it two point} equilibrium spatio-temporal correlations of the two conserved fields, namely, norm and energy. In particular, the high temperature regime is characterized by diffusive spreading of the correlations with zero sound velocity. On the contrary, the low temperature regime shows super-diffusive spreading of the correlations which travel ballistically with the speed of sound. This is rooted in the existence of an additional almost conserved field (total phase difference). Consequently, in this temperature regime, the density–density correlations have symmetrically located sound peaks travelling ballistically in opposite directions and broadening as $t^{z}$ with $z=2/3$ \cite{Kulkarni_2013,Kulkarni_2015,Mendl_2015}. Therefore, the dynamical critical phenomena falls under the Kardar-Parisi-Zhang (KPZ) universality class. Not only the exponent but the functional form of the correlations also matches the Pr\"{a}hofer-Spohn scaling function \cite{Prahofer_2004}. This mapping of the DNLS system to the KPZ universality class has been thoroughly discussed in Ref.~\onlinecite{Kulkarni_2013,Mendl_2015,Kulkarni_2015}. In addition to these two sound modes, one has a central (non-moving) heat peak that broadens as $t^{3/5}$ with a L\'{e}vy $\frac{5}{3}$ shape function \cite{Mendl_2015}.  
Interestingly, an  almost integrable structure emerges in the ultra-low temperature regime manifesting as ballistic broadening of all correlations \cite{Mendl_2015,Iubini_2012}. 

While the {\it two point} correlations have proven to be remarkable diagnostics of the three regimes \cite{Mendl_2015}, it would be interesting 
to ask if this information about the existence of different dynamical regimes, can be extracted through the study of even simpler {\it one point} macroscopic thermodynamic observables, like temperature, average energy density, average norm density etc. We successfully address this question in this paper by looking at the relationship between energy density and temperature. We also probe the three dynamical regimes by analyzing the emergence and disappearance of an additional conserved quantity through phase slip events \cite{Das_2020}. We show that the different regimes can be investigated through the lens of higher order correlations such as the classical analogue of out-of-time-ordered correlators \cite{Das_2018,Bilitewski_2018,Kumar_2019,Chatterjee_2020,Ruidas_2020,Bhanu_2020,Bilitewski_2020}. 


In this paper, we consider the one-dimensional DNLS in equilibrium. To probe the different dynamical regimes using one point thermodynamic observables, we investigate the system in grand canonical ensemble. Notably, the grand canonical ensemble is implemented by connecting the DNLS to two Langevin thermostats  at same temperature and chemical potential at the chain ends \cite{Iubini_2013}. 
Below, we briefly summarize our main observations.

(i) We put forward a fascinating yet simple diagnostic for the three distinct dynamical regimes of DNLS in equilibrium. We numerically find the power-law relationship $T=c\,\epsilon^\alpha$, where $T$ is the temperature, $\epsilon$ is the average energy density and $c$ is a constant. The exponent $\alpha$ serves as a demarcator of the three regimes. More precisely, we observe that $\alpha>1$ for high temperature regime, $\alpha<1$ for low temperature regime and $\alpha=1$ for ultra-low temperature regime (Table.~\ref{tab:table}). 
We note that the temperature at which the minimum of the ratio $r(\epsilon) = \frac{T}{\epsilon}$ occurs defines  the crossover temperature $T_{h-l}$ between the high temperature and low temperature regimes. The crossover temperature ($T_{h-l}$) obtained by this method compares extremely well with the 
criterion for crossover temperature proposed in Ref.~\onlinecite{Mendl_2015}. As we decrease temperature further, $r(\epsilon)$ starts increasing and saturates to a constant below an ultra-low temperature $T_{l-ul}$ as expected for harmonic chains. 

(ii) The DNLS has two conserved quantities, namely the total energy and the total mass (norm). Interestingly, we observe the emergence of an additional almost conserved quantity (total phase difference) in the low temperature regime. This distinguishes the low temperature regime from the high temperature regime where this third conservation law does not hold. To understand this, we probe the system using the concept of dynamically activated processes that lead to discontinuous jumps in the phase differences, known as phase slips. We find that the frequency of these phase slip events increase exponentially as one enters the high temperature regime, thereby resulting in the violation of the additional conservation law. Remarkably, the significant difference in the activation energies required for the phase slip events, demarcates the low temperature regime from the high temperature regime. In the ultra-low temperature regime, we observe no phase slip events even for extremely long times.

(iii) The DNLS is known to be generically non-integrable \cite{Ablowitz_2004} and chaotic in nature. However, it shows almost integrable 
features at very low temperatures \cite{Iubini_2013}. To investigate this in detail, we study chaos in the different dynamical regimes of the DNLS. As tools, we have used the classical out-of-time-ordered correlator (OTOC) and related observables, namely the butterfly speed and Lyapunov exponents \cite{Das_2018,Bilitewski_2018,Kumar_2019,Chatterjee_2020,Ruidas_2020,Bhanu_2020,Bilitewski_2020}. In particular, the butterfly speed (measuring the speed of spatial propagation of chaos) exhibits intriguing non-monotonic behavior with temperature. Furthermore, the crossover temperature $T_{l-ul}$ (between low and ultra-low temperature regimes) is interestingly given by the temperature at which minimum of the butterfly speed occurs. The Lyapunov exponent, on the other hand, follows a monotonically increasing power-law behavior with an exponent $\gamma$. Remarkably, the value of $\gamma$ changes considerably along the crossovers between different dynamical regimes. The space-time heat-maps of the OTOC displays visibly striking differences between the ultra-low temperature regime (oscillatory structures in space-time inside the light-cone) and the low/high temperature regimes (exponential growth inside the light-cone with oscillatory structures absent).

The paper is organized as follows. In section \ref{model}, we describe the model and discuss in detail how to set up the system in grand canonical ensemble. The numerical results concerning the temperature-energy relationship distinguishing the three dynamical regimes, are presented in section \ref{sec:virial}. In section \ref{sec:phaseslip}, we analyze the different dynamical regimes through the emergence and disappearance of an additional almost conserved quantity. In section \ref{sec:otoc}, we probe the chaotic nature of DNLS in the three dynamical regimes using OTOC, butterfly speed and Lyapunov exponent. We conclude with a  brief summary of our observations and future directions in section \ref{summary}. The details of the numerical procedures used here and the relevant error analysis are presented in Appendix \ref{app:numerics} and Appendix \ref{app:error} respectively. 

\section{Model and observables}
\label{model}
The Hamiltonian of a discrete nonlinear Schr\"{o}dinger chain (DNLS) defined on a one dimensional lattice with $N$ sites is given by 
\be
H=\sum_{j=1}^{N} \left(|\psi_{j+1}-\psi_{j}|^2 + \frac{g}{2}|\psi_j|^4 \right).
\label{eq:ham}
\ee
Here   $\psi_j$ $(j=1,2 \dots N)$ is a complex valued field and  $g>0$ is the defocusing nonlinearity parameter  \cite{Ablowitz_2004,Hasegawa_1973}. 
This system has two conserved quantities, total energy $E$ and the total `mass' $A$ \cite{Iubini_2012}. 
The  mass conservation is equivalent to the normalization condition of the complex field $\psi_j$ which is given by 
\be
A=\sum_{j=1}^{N} |\psi_j|^2.
\label{eq:mass}
\ee
Here $|\psi_j|^2$ can be interpreted as the local mass density associated with the site $j$. 
In case of periodic boundary conditions ($\psi_{N+i}=\psi_i$), it is easy to see that the Hamiltonian in Eq.~\eqref{eq:ham} can be re-written as
\be
H= 2 A\,+\,\sum_{j=1}^{N} \left[ -\left(\psi_{j+1}\psi_j^\ast + \psi_{j+1}^\ast\psi_j\right)+ \frac{g}{2}|\psi_j|^4 \right].
\label{eq:ham1}
\ee
One should note that the minus sign in front of the hopping term is irrelevant due to the symmetry associated with a suitable gauge transformation of the form $\psi_j\rightarrow e^{\mathrm{i}\pi j}\psi_j$. Using this gauge transformation, the Hamiltonian in Eq.~(\ref{eq:ham1}) becomes
\be
H= 2 A\,+\,\sum_{j=1}^{N} \left[ \left(\psi_{j+1}\psi_j^\ast + \psi_{j+1}^\ast\psi_j\right)+ \frac{g}{2}|\psi_j|^4 \right].
\label{eq:ham2}
\ee
In order to bring out the interesting non-separable structure of the DNLS Hamiltonian, a pair of canonically conjugate variables $(q_j,p_j)$ can be introduced as 
\be
\psi_j= \frac{1}{\sqrt{2}}(q_j + \mathrm{i} p_j), \,\,\,\,\,
\psi_j^\ast= \frac{1}{\sqrt{2}}(q_j - \mathrm{i} p_j),
\label{eq:qp}
\ee
where both $q_j$ and $p_j$ are real valued variables $\forall j$. Accordingly, the two conserved quantities, the energy given by the Hamiltonian  in Eq.~(\ref{eq:ham2}) and mass given in Eq.~(\ref{eq:mass}) take the following forms
\begin{align}
\begin{split}
H &= 2 A\,+ H_{\text{DNLS}}, \cr
H_{\text{DNLS}} &= \,\sum_{j=1}^{N} \left[ \left(q_{j+1}q_j + p_{j+1}  p_j\right)+ \frac{g}{8}\left(q_j^2+p_j^2\right)^2 \right], \cr
A &= \frac{1}{2}\sum_{j=1}^{N} \left(q_j^2+p_j^2\right).
\end{split}
\label{eq:hamqp}
\end{align}
It is evident from Eq. (\ref{eq:hamqp}) that the quartic on-site nonlinear term represents the interaction between $q$ and $p$ degrees of freedom, whereas the hopping terms imitate the coupling between the degrees of freedom at nearest-neighbor sites.
Importantly, we should note how the introduction of the canonical co-ordinates $(q_i,p_i)$ clearly exhibits the non-separable structure of the DNLS Hamiltonian. More precisely, from the expression of the Hamiltonian in Eq. (\ref{eq:hamqp}), we observe that $q_i$-s and $p_i$-s are not the usual positions and momenta. Consequently, $H$ is not in the usual {\it separable} sum form of kinetic energy and potential energy. Rather, the Hamiltonian has a non-trivial {\it non-separable}  form (symmetric under the exchange $q_i\leftrightarrow p_i$) giving rise to intriguing dynamical features.

In this paper, we consider the 1D DNLS in a grand canonical ensemble.  For this purpose, we connect the chain with two Langevin thermostats of same temperature $T$ and same chemical potential $\mu$ at it's two ends ($j=1$ and $j=N$) \cite{Iubini_2013}. 
To model DNLS in equilibrium, both the thermostats are kept at same temperature $T$ and same chemical potential $\mu$. 
The Langevin dynamics should be chosen in such a way that the system finally relaxes to the grand canonical equilibrium distribution
\bea
P(\left\lbrace p_i,q_i \right\rbrace) 
&=&\frac{e^{-\beta(H_{\mathrm{DNLS}}-\mu A)}}{Z}=\frac{e^{-\beta H_{\mu}}}{Z}\cr
H_\mu &=& H_{\text{DNLS}} - \mu A,
\label{eq:hmu}
\eea
where $H_{\text{DNLS}}$ and $A$ are given in Eq.~(\ref{eq:hamqp}) and $Z$ is the partition function. 
%
In Ref.~\onlinecite{Iubini_2013}, it has been demonstrated that the following Langevin equations take the system to the above equilibrium state (Eq.~\ref{eq:hmu}).
\begin{align}
\begin{split}
\dot{q_1}&=\frac{\partial H_{\mathrm{DNLS}}}{\partial p_1}-\gamma \frac{\partial H_\mu}{\partial q_1}+\sqrt{2\gamma T}\, \xi_{1}'(t)\\
 \dot{p_1}&=-\frac{\partial H_{\mathrm{DNLS}}}{\partial q_1}-\gamma \frac{\partial H_\mu}{\partial p_1}+\sqrt{2\gamma T}\, \xi_{1}''(t)\\
 \dot{q_j}&=\frac{\partial H_{\mathrm{DNLS}}}{\partial p_j}\;\;\;\;\;\; \mathrm{for}\;\; j=2 \dots N-1\\
 \dot{p_j}&=-\frac{\partial H_{\mathrm{DNLS}}}{\partial q_j}\;\;\;\;\; \;\;\mathrm{for}\;\; j=2 \dots N-1 \\
 \dot{q_N}&=\frac{\partial H_{\mathrm{DNLS}}}{\partial p_N}-\gamma \frac{\partial H_\mu}{\partial q_N}+\sqrt{2\gamma T}\, \xi_{N}'(t)\\
 \dot{p_N}&=-\frac{\partial H_{\mathrm{DNLS}}}{\partial q_N}-\gamma \frac{\partial H_\mu}{\partial p_N}+\sqrt{2\gamma T}\, \xi_{N}''(t)
\end{split}
 \label{eq:eqm}
\end{align}
where $\gamma>0$ is the coupling strength between the system and reservoirs. $\xi_{1}',\xi_{1}'',\xi_{N}',\xi_{N}''$ are Gaussian white noises each of which is delta correlated i.e. $\langle\xi(t)\xi(s)\rangle=\delta(t-s)$ and has zero mean. The explicit expressions for the equations of motion in 
Eq.~(\ref{eq:eqm}) and the numerical methods used for the corresponding numerical integration are discussed in detail in Appendix \ref{app:numerics}.





\begin{figure*}
 \centering
  \subfigure[]{\includegraphics[scale=0.35]{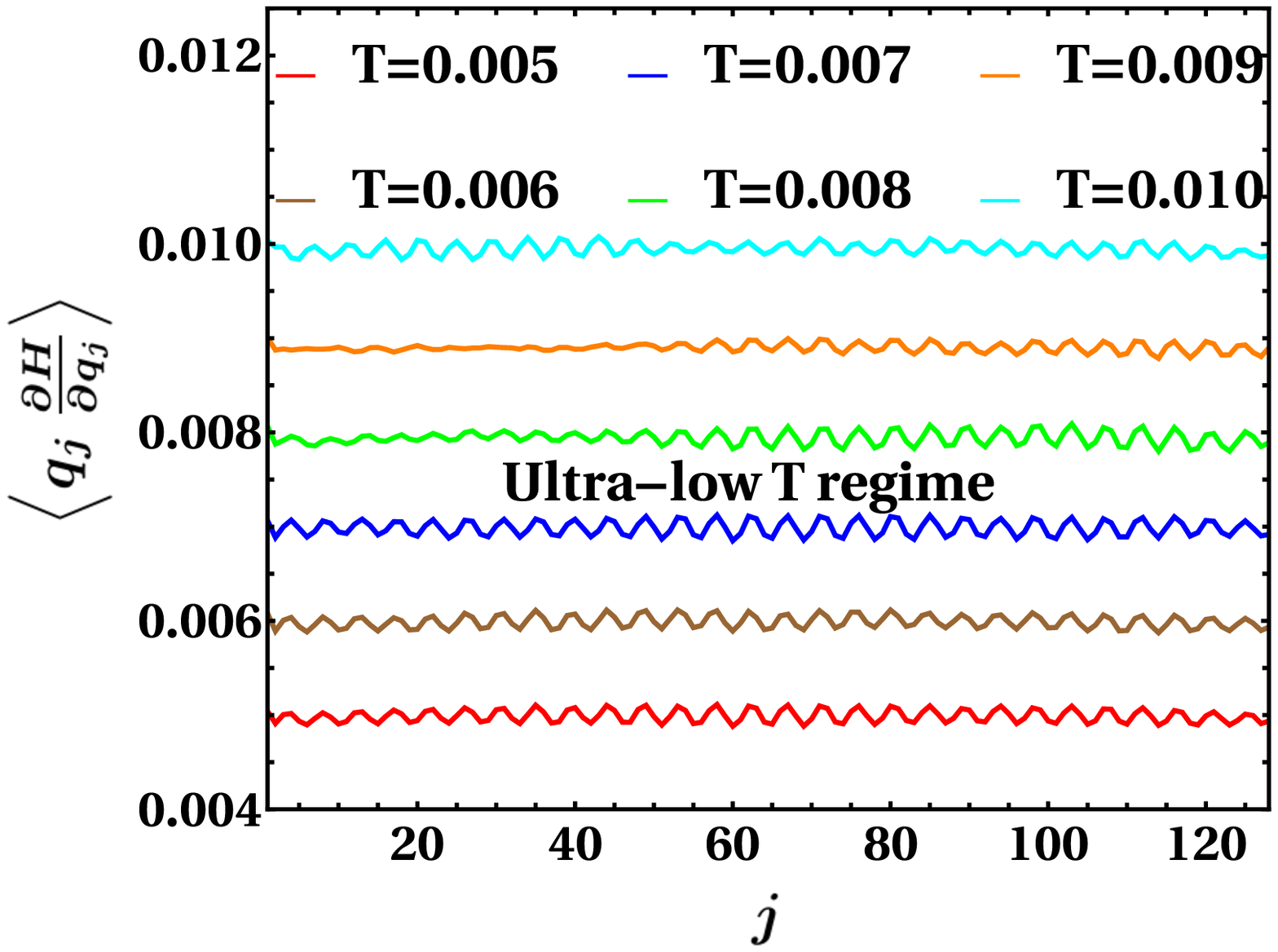}}\hfill
  \subfigure[]{\includegraphics[scale=0.35]{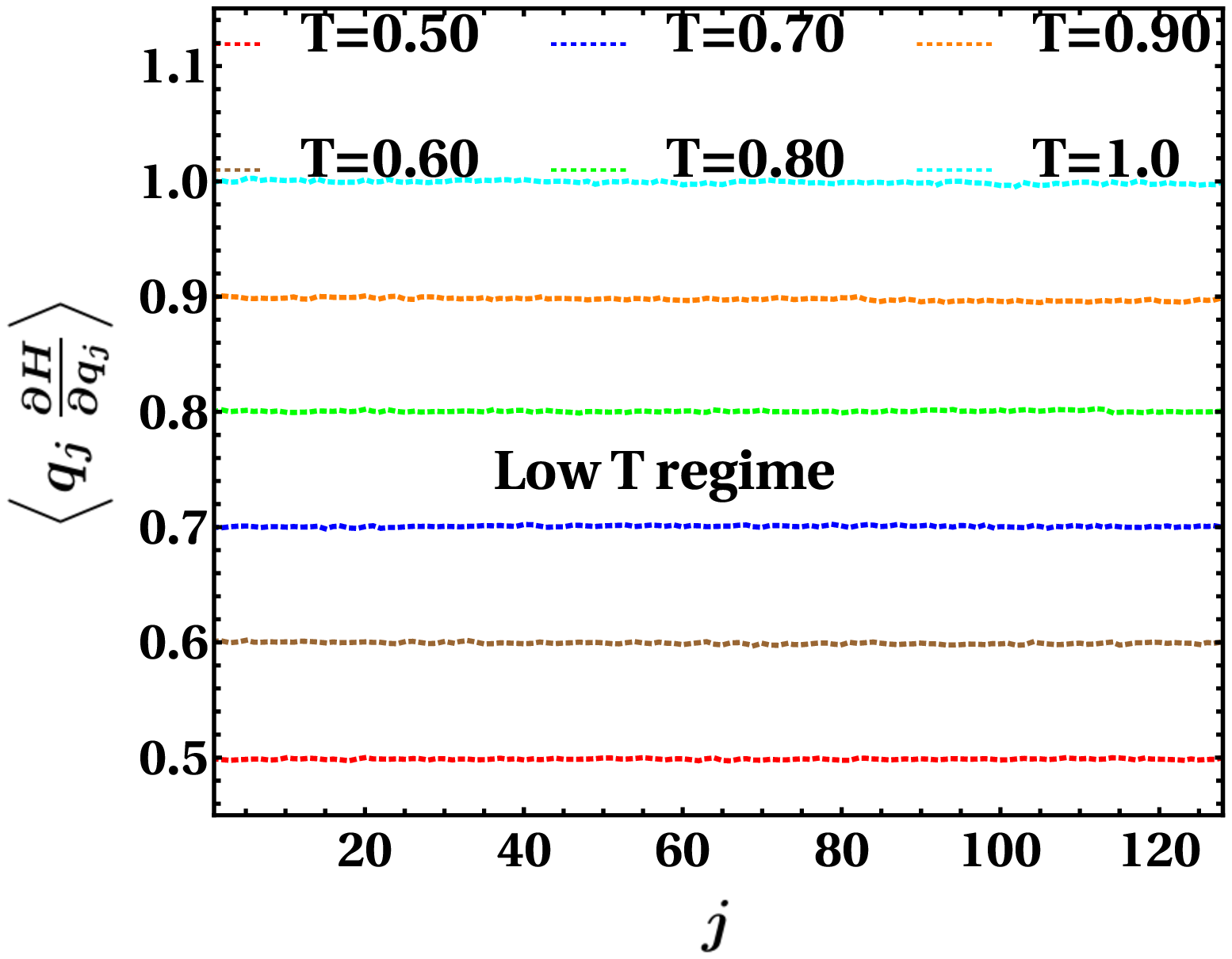}}\hfill
  \subfigure[]{\includegraphics[scale=0.35]{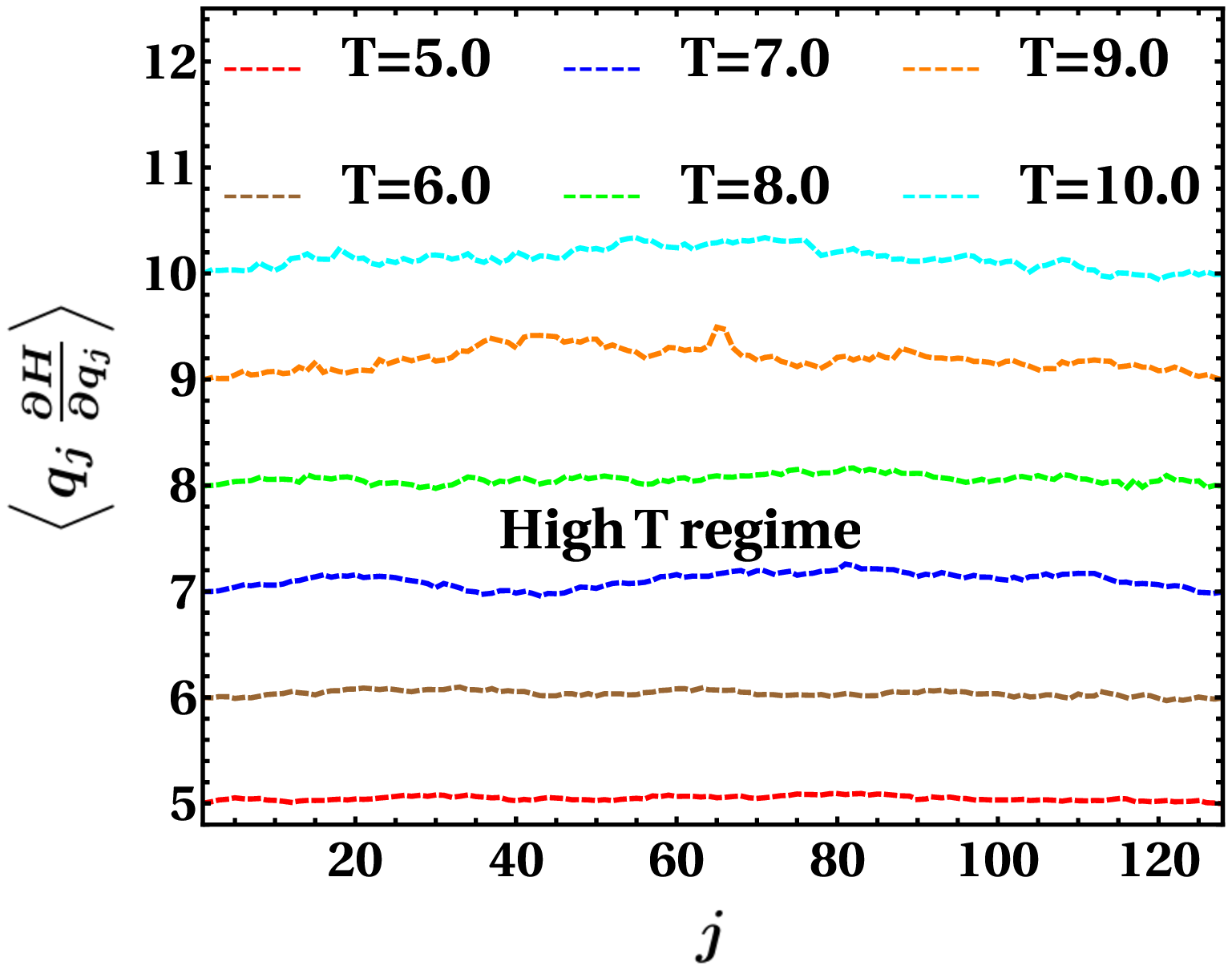}}\\
  \hspace*{0.3 cm}\subfigure[]{\includegraphics[scale=0.35]{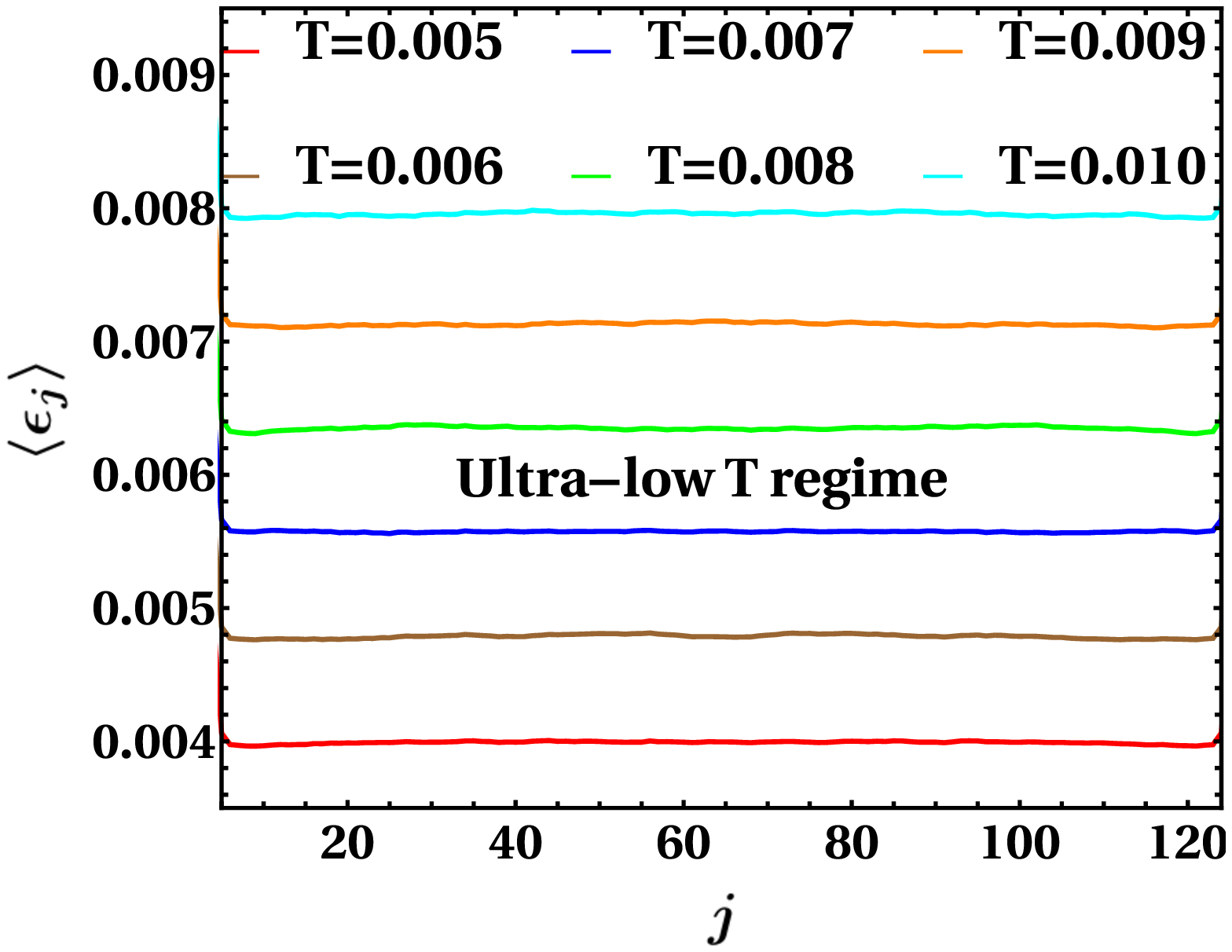}}\hfill
  \subfigure[]{\includegraphics[scale=0.35]{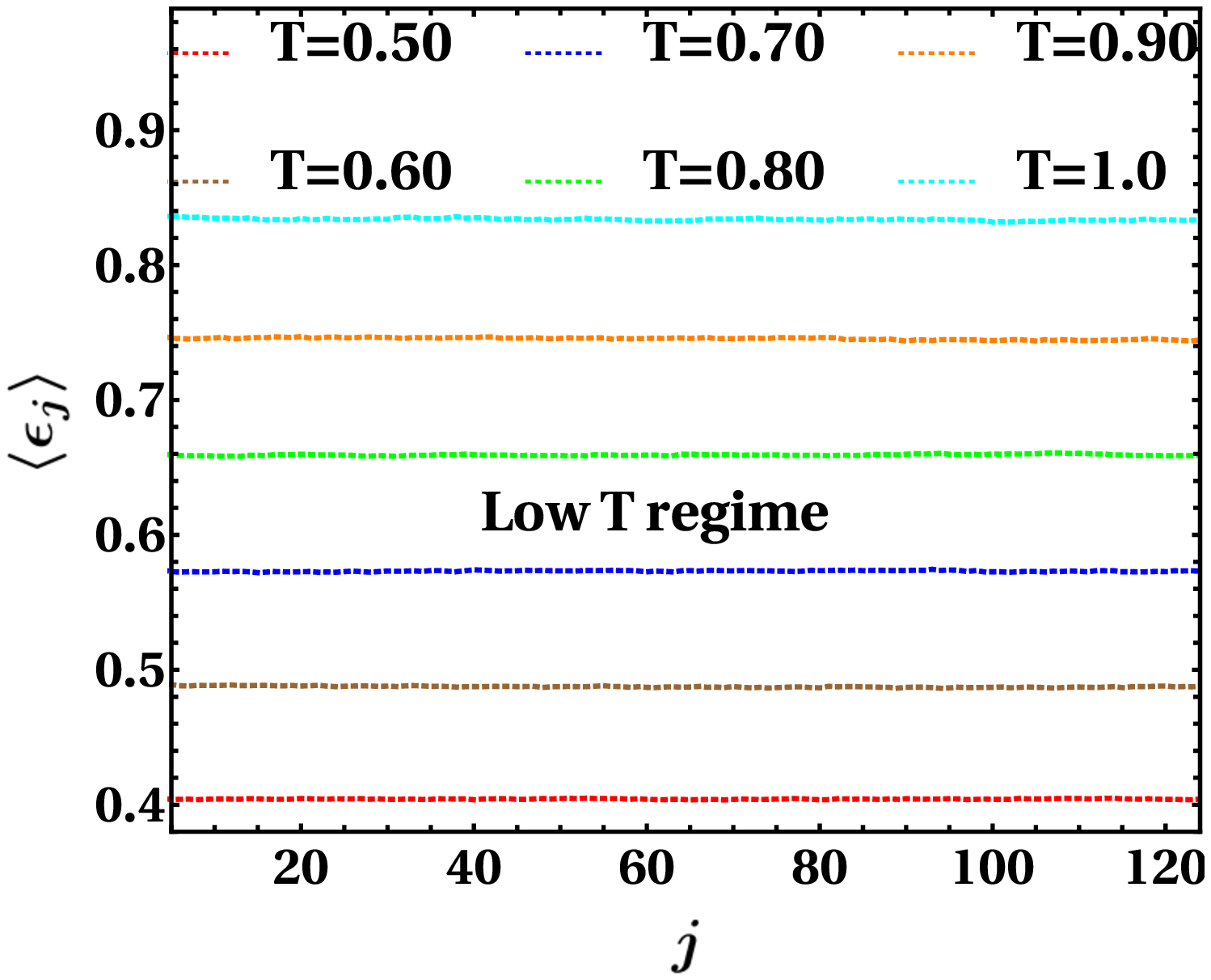}}\hfill
  \subfigure[]{\includegraphics[scale=0.35]{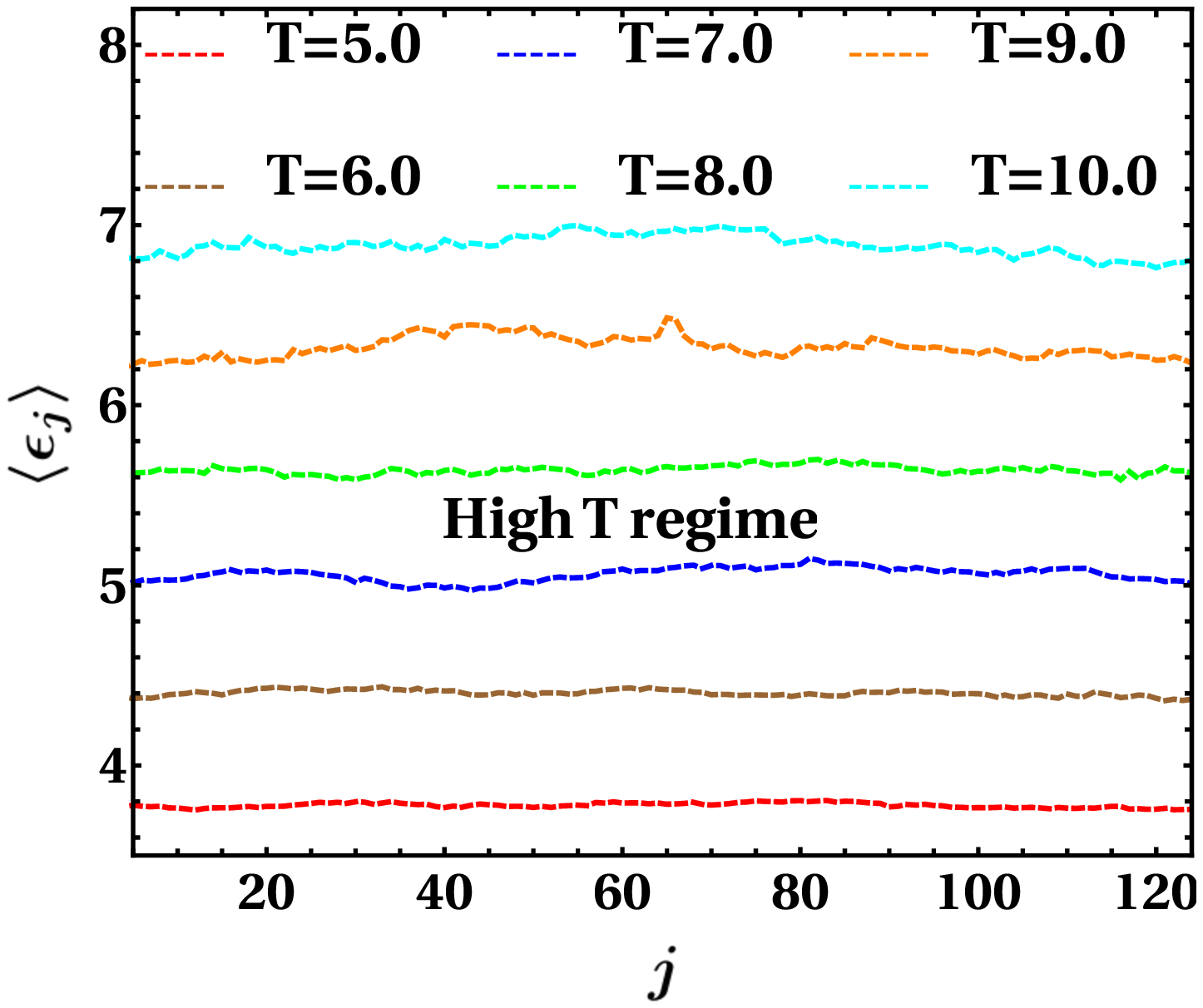}}
  \caption{Figures (a), (b) and (c) in the upper panel show that the long time averages of the virial observable (Eq. \ref{eq:virial}) converge very well to the temperatures of the Langevin thermostats, in ultra-low temperature, low temperature  and high temperature regime respectively. The equilibration is also ensured by the flat spatial profiles of the average energy density $\langle\epsilon_j\rangle$ (Table \ref{tab:observable}) presented in the lower panel figures (d), (e) and (f). }
\label{fig:vir_can}
\end{figure*}

The main goal of this work is to show that even relationships between simple one point thermodynamic observables, average energy density ($\epsilon$) and temperature ($T$), clearly exhibits the existence of three different dynamical regimes of DNLS in equilibrium, investigated earlier in Ref.~\onlinecite{Mendl_2015} using higher order two point equilibrium spatio-temporal correlations. Also, we analyze the distinction between the different regimes through the emergence and disappearance of an additional almost conserved quantity (total phase difference) using the concept of phase slip events \cite{Das_2020}. In addition, we would like to see the signatures of the different dynamical regimes in higher point correlations of the microscopic degrees of freedom, through OTOC, butterfly speed and Lyapunov exponent  \cite{Das_2018,Bilitewski_2018,Kumar_2019,Chatterjee_2020,Ruidas_2020,Bhanu_2020,Bilitewski_2020}. In this connection, we numerically compute the following quantities - temperature $T$ using a generalized virial theorem,  average energy density ($\epsilon$) and average mass density ($\rho$) [see section \ref{sec:virial}], OTOC [$D_x(i,t)$] (see section \ref{sec:otoc}).  This is done using Eq.~(\ref{eq:eqm}) [see Appendix~\ref{app:numerics} for details]. The various observables analyzed in this work are explicitly given in Table.~\ref{tab:observable}.
\begin{table*}[ht]
\begin{center}
\begin{tabular}{|p{0.25\textwidth}|p{0.50\textwidth}|}
\hline
\textbf{Observable} & \textbf{Expression}\\
\hline
Temperature ($T$) & $T=\left\langle x_j \frac{\partial H_\mu}{\partial x_j}\right\rangle$, \hspace*{0.1 cm} $x=q,p$ \\
\hline
Average energy  density ($\epsilon$) & $\epsilon=\langle\epsilon_j\rangle$\\
~~ &  $\epsilon_j=\frac{1}{4m}[\left(q_{j-1}+q_{j+1}\right)q_j+ \left(p_{j-1}+p_{j+1}\right)p_j] + \frac{g}{8}\left(q_j^2+p_j^2\right)^2$\\
\hline
Average mass density ($\rho$) & $\rho=\langle\rho_j\rangle$, $\rho_j= \frac{1}{2}\left(q_j^2+p_j^2\right)$ \\
\hline\hline
OTOC $\left[D_x(j,t)\right]$ & $D_x(j,t)=\left\langle \left|\frac{\delta x_j(t)}{\delta x_k(0)}\right|\right\rangle_{ic}$, \hspace*{0.1 cm} $x=q,p$ \\
\hline
Lyapunov exponent [$\lambda_x(j)$] & $\lambda_x(j) =\mathrm{lim}_{t\rightarrow \infty}\left\langle \frac{1}{t} \mathrm{ln}~D_x(j,t)\right\rangle_{ic}$ \\
\hline
Butterfly speed ($v_b$) & $\frac{1}{t}\left\langle\sum_{j=1}^{N}\Theta\left(\frac{\delta x_j(t)}{\delta x_k(0)}-1\right)\right\rangle_{ic}$, $x=q,p$ \\
\hline
\end{tabular}
\caption{The table contains the list of observables and their corresponding mathematical expressions, that we compute numerically for the DNLS. The first three rows are concerning one point correlators and the last three rows are concerning higher point correlators. Here $\langle.\rangle$ denotes time average whereas $\langle.\rangle_{ic}$ denotes average over initial conditions in equilibrium.}
\label{tab:observable}
\end{center}
\end{table*}

In the subsequent sections, using the grand canonical set up described here, we compute these quantities numerically and discuss how their behaviour distinguishes the three temperature regimes mentioned above. 

\section{Dynamical regimes}
\label{sec:virial}
In this section, we would like to present and analyze
the numerical results on temperature ($T$), average energy density ($\epsilon$) and the relation between them. We show how this relation between one point thermodynamic functions demarcates the different dynamical regimes - the ultra-low temperature, low temperature and high temperature regimes. As mentioned earlier, the existence of these three regimes has been reported recently in Ref.~\onlinecite{Mendl_2015}, but by means of higher order two point  equilibrium spatio-temporal correlations of the conserved quantities $\epsilon$ and $\rho$. It is pertinent to mention that the grand canonical (Eq.~\ref{eq:eqm}) set up with Langevin thermostats thermalize the DNLS system \cite{Iubini_2013}. While it is relatively easy and conventional to thermalize a Hamiltonian with separable structure \cite{Gardiner_2004} in numerical simulations, thermalizing a non-separable Hamiltonian such as DNLS is far from obvious \cite{Iubini_2013} (see Appendix \ref{app:numerics} for details). Therefore, naturally, this demands for a rigorous and careful check for thermalization  in the wide range of temperatures starting from ultra-low temperature regime up to the high temperature regime. To study thermalization, we take aid of the generalized virial theorem stated below.

\begin{figure*}
  \centering
  \subfigure[]{\includegraphics[scale=0.38]{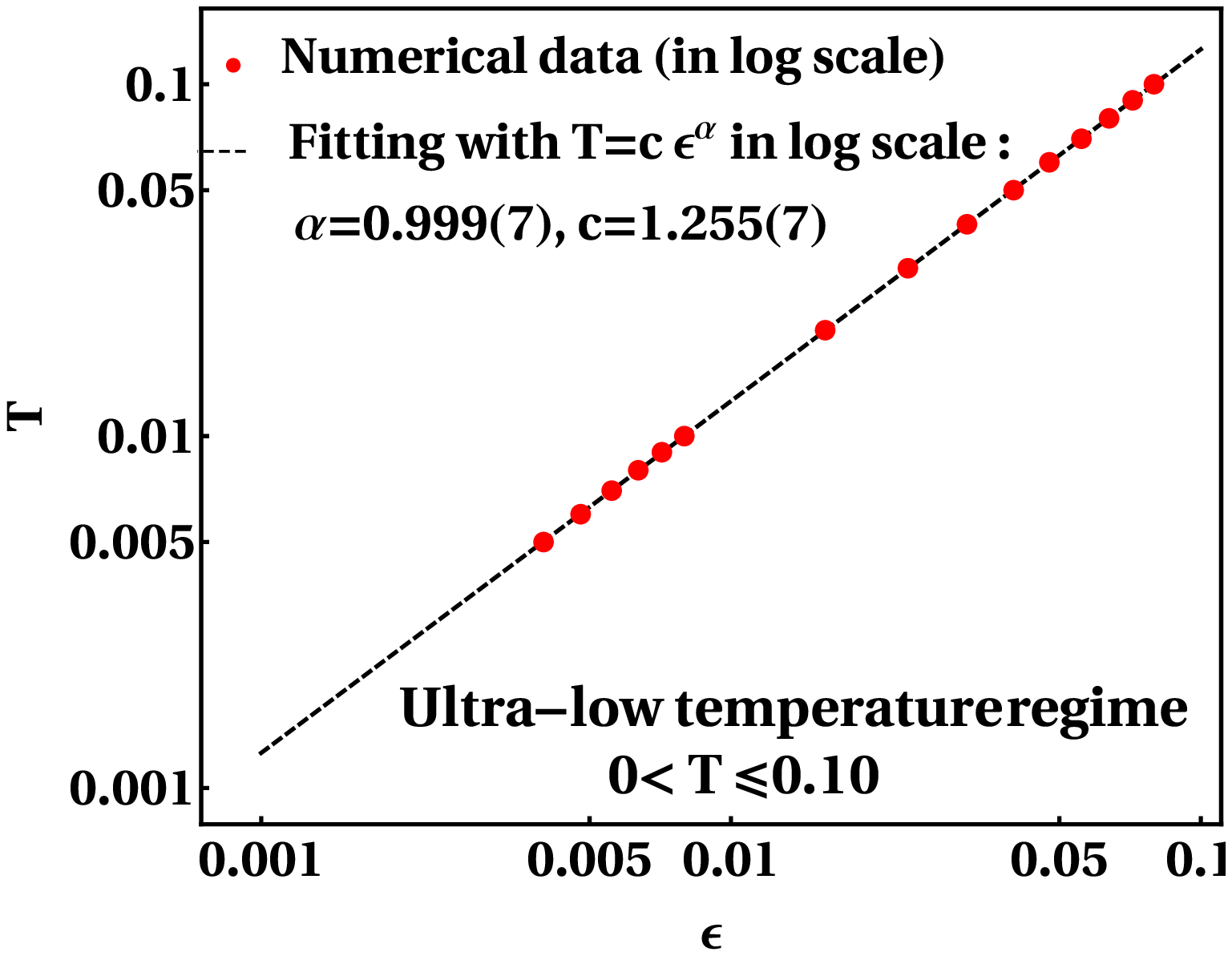}}\hfill
  \subfigure[]{\includegraphics[scale=0.38]{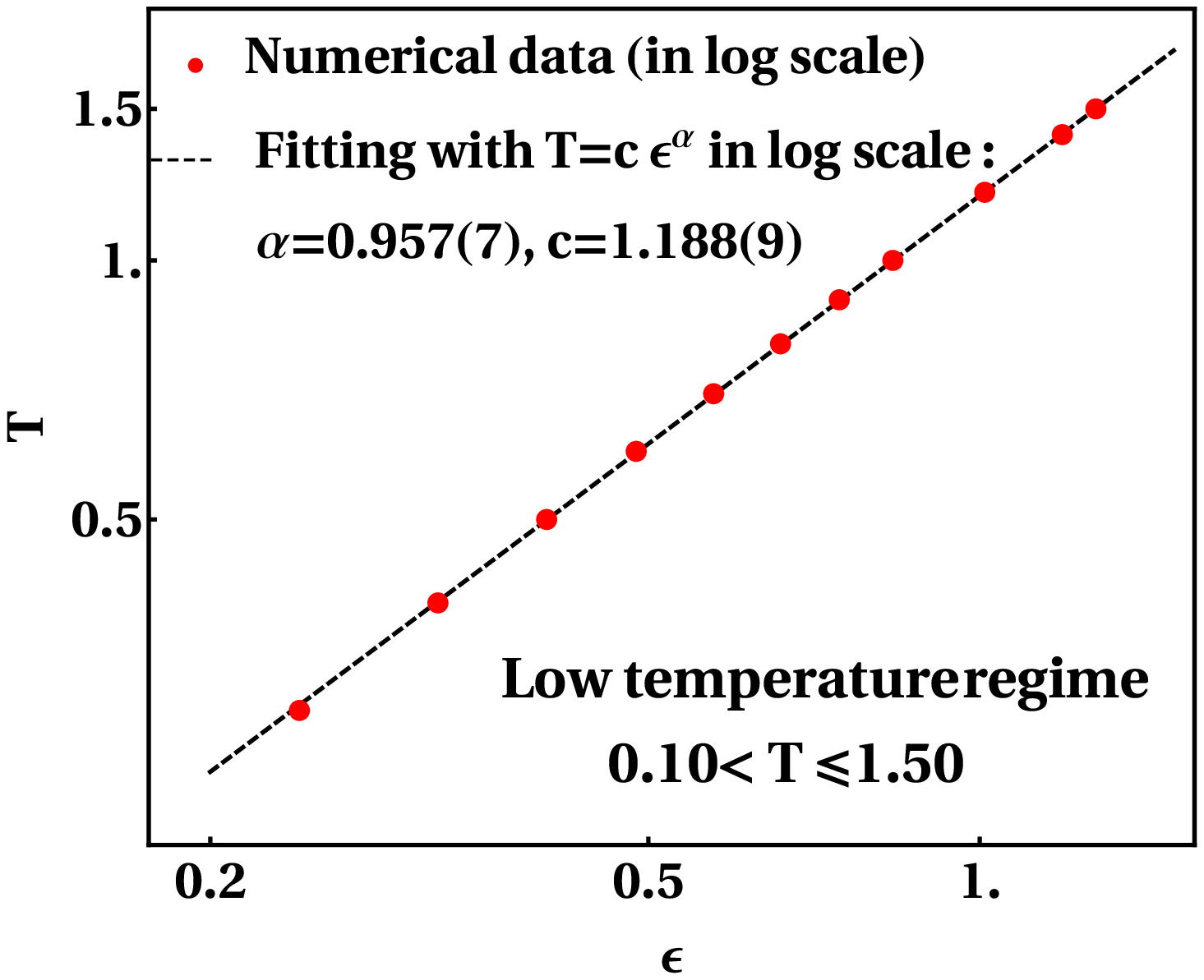}}\hfill
  \subfigure[]{\includegraphics[scale=0.38]{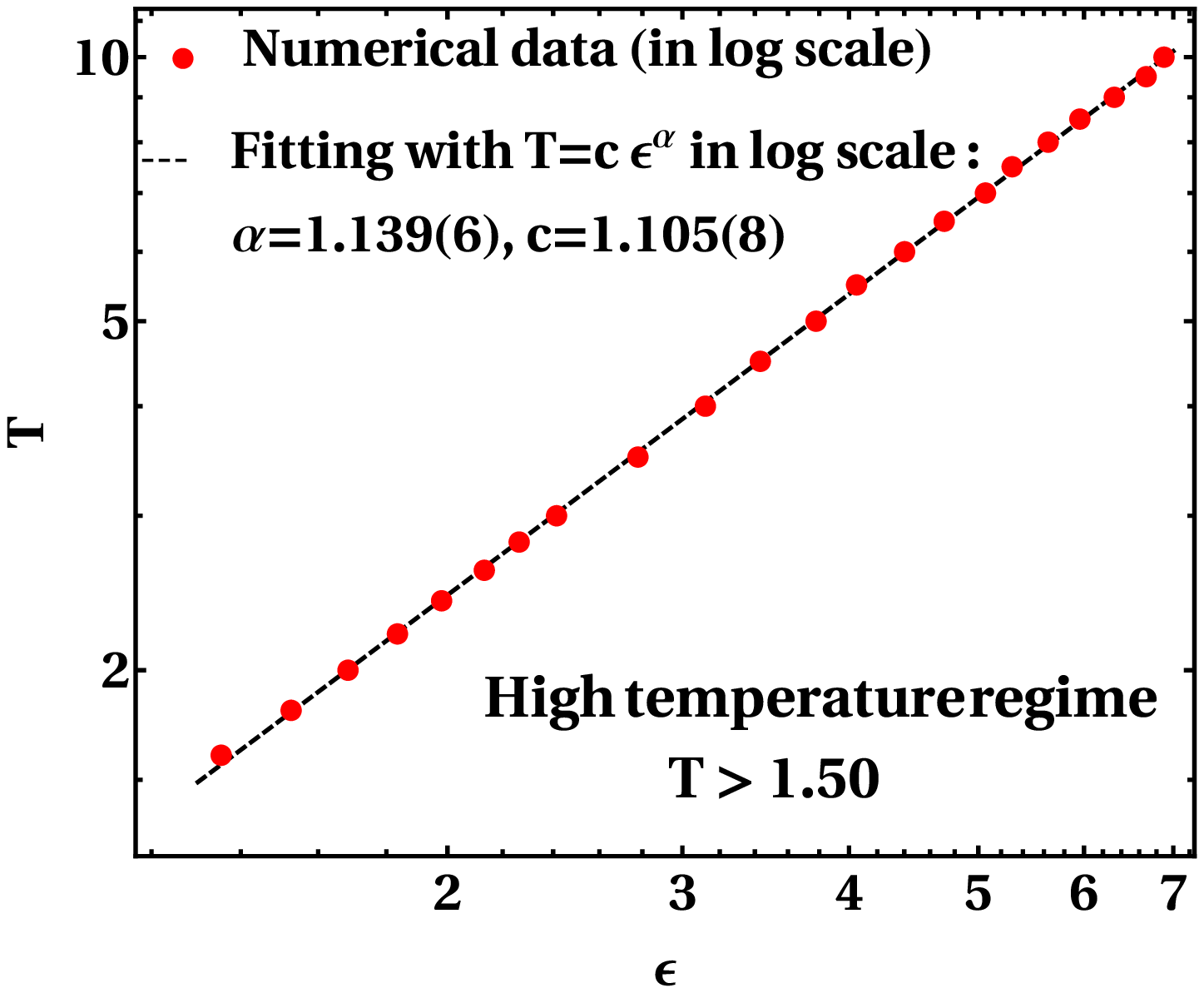}}\\
  \caption{In this figure, we present the numerical data for $T$ versus $\epsilon$ in log-log scale. Interestingly, we  observe the existence of three different temperature regimes of the DNLS, clearly  demarcated by the value of the exponent $\alpha$ (Eq. \ref{eq:tphi}). The sub-figures (a), (b) and (c) represent ultra-low temperature, low temperature and high temperature regime characterized by $\alpha=1$, $\alpha<1$ and $\alpha>1$ respectively. Parameters used are $g=2.0$, $\mu=2.0$ and the chain length is $N=128$. }
\label{fig:difft}
\end{figure*}

\subsection{Generalized virial theorem}


The generalized virial theorem \cite{Pathria_1986} states how the macroscopic temperature $T$ of the system is related to the microscopic degrees of freedom $q_j$-s (or equivalently $p_j$-s). It is given by
\be
\left\langle q_j \frac{\partial H_\mu(\left\lbrace q_j,p_j\right\rbrace)}{\partial q_j}\right\rangle = k_B T,
\label{eq:virial}
\ee
where $H_\mu(\left\lbrace q_j,p_j\right\rbrace)$ is the Hamiltonian of the system. For DNLS, the Hamiltonian $H_\mu$ is given in Eq.~(\ref{eq:hmu}) and $\langle.\rangle$ denotes time average over equilibrium state. We set Boltzmann constant $k_B=1$ throughout the manuscript. Through involved numerical integration of the equations of motion (Eq.~\ref{eq:eqm}), we compute the virial observable $\left\langle q_j \frac{\partial H_\mu}{\partial q_j}\right\rangle$ and observe that, indeed, the time average of this quantity converges very well to the temperature $T$ of the Langevin thermostats. In the upper panel of Fig.~\ref{fig:vir_can}, we verify Eq.~(\ref{eq:virial}) for three different temperature regimes of the thermostats - (a) ultra-low temperature, (b) low temperature and (c) high temperature. 
The equilibration in the DNLS is further ensured by investigating spatial profiles of the energy density $\langle\epsilon_j\rangle$ (Table \ref{tab:observable}), which is plotted in the lower panel of Fig.~\ref{fig:vir_can} for the respective temperature regimes. At sufficiently high temperatures, we notice some spatial fluctuations of the virial observables around the mean [Fig. \ref{fig:vir_can}(c) and Fig. \ref{fig:vir_can}(f)]. It is pertinent to mention that the extents of these spatial fluctuations are small as supported by the error analysis in Appendix \ref{app:error} (see Fig.~\ref{fig:dphi}).


\subsection{$T-\epsilon$ relation}

Having computed $T$ and $\epsilon$ in the previous section, one naturally wonders about the intricate relationship between them. In simple linear systems (for e.g., Harmonic chain), these two are proportional to each other. However in nonlinear systems, this relationship is far from obvious which is what we investigate here. We demonstrate below that this relation becomes instrumental to identify the three temperature regimes mentioned earlier.



For usual separable Hamiltonian with potential energy being a homogeneous function of degree $\eta$, the 
generalized virial theorem in Eq.~(\ref{eq:virial}) results in the following $T-\epsilon$ relationship \cite{Howard_2005}
\be
T= \frac{2 \eta}{\eta +2}\epsilon=r\,\epsilon.
\label{eq:r}
\ee

For example, for 
a coupled Harmonic chain with Hamiltonian $H=\sum_{j=1}^{N}\left[\frac{p_j^2}{2m}+\frac{q_j^2}{2m}+\left(q_{j+1}-q_j\right)^2\right]$ ($q_j, p_j$ being usual position and momentum respectively), the potential energy is a homogeneous function of degree $\eta=2$. Consequently, Eq.~(\ref{eq:r}) implies that $r=1$ and $T=\epsilon$ for the coupled Harmonic chain. Whereas if one considers a different interaction with $\eta=4$ i.e. a coupled quartic chain with Hamiltonian $H=\sum_{j=1}^{N}\left[\frac{p_j^2}{2m}+\frac{q_j^4}{2m}+\left(q_{j+1}-q_j\right)^4\right]$, Eq.~(\ref{eq:r}) directly says that we have a different $T-\epsilon$ relation of the form $T=\frac{4}{3}\epsilon$ with $r=4/3$. 
It is far from obvious how Eq.~(\ref{eq:r}) gets generalized for the non-separable Hamiltonian such as DNLS [Eq.~(\ref{eq:hamqp})] which is also inhomogeneous. 
To investigate this, 
we plot $T$ versus $\epsilon$ in Fig.~\ref{fig:difft} (in the three different temperature regimes) and try to fit the corresponding data points using a power law of the form
\be
T=c\,\epsilon^\alpha.
\label{eq:tphi}
\ee

We find that the DNLS indeed has three different temperature regimes clearly demarcated from each other by the value of the exponent $\alpha$. Interestingly, from the sub-figures (a), (b) and (c) in Fig.~\ref{fig:difft}, we observe that $\alpha = 1$, $\alpha<1$ and $\alpha>1$ in these three different regimes, which we call as ultra-low temperature regime, low temperature regime and high temperature regime respectively, following the nomenclature used in Ref.~\onlinecite{Mendl_2015}. The crossover temperatures between high and low temperature regimes and that of the low and ultra-low temperature regimes, would be referred to as $T_{h-l}$ and $T_{l-ul}$ respectively. More elaborately, for $0<T< T_{l-ul}$, the DNLS is in ultra-low temperature regime characterized by $\alpha=1$ [Fig.~\ref{fig:difft}(a)]. Clearly, at ultra-low temperatures, $\alpha=1$ infers an almost linear $T-\epsilon$ relation similar to separable Hamiltonian systems with homogeneous potentials as discussed in Eq.~(\ref{eq:r}). In contrast, the non-trivial nature of the DNLS Hamiltonian becomes apparent from the strongly nonlinear relation between $T$ and $\epsilon$ in the high temperature regime $T> T_{h-l}$ where $\alpha>1$. More intriguingly, in  the intermediate low temperature regime where $T_{l-ul}<T<T_{h-l}$, the $T-\epsilon$ relation is still nonlinear but with $\alpha<1$. For the particular example shown in Fig.~\ref{fig:difft} with specific parameter values, the differences between the values of $\alpha$ in different temperature regimes are small. Therefore, we calculate the corresponding error bars systematically in Appendix \ref{app:error} and show that the error bars are indeed negligibly small compared to the differences in the $\alpha$ values (Fig.~\ref{fig:dalpha}). So, we conclude that the exponent $\alpha$ (appearing in the relation Eq. (\ref{eq:tphi}) between one point macroscopic observables $T$ and $\epsilon$), acts as a remarkable identifier of the different dynamical regimes of the DNLS. 

Based on Eq. (\ref{eq:tphi}) and the corresponding observations from Fig.~\ref{fig:difft}, it seems reasonable to define an empirical specific heat, $C(T)$ as 
\be
C(T)=\frac{\partial \epsilon}{\partial T}=\frac{1}{\alpha\,c^{\frac{1}{\alpha}}}~T^{\frac{1-\alpha}{\alpha}}.
\label{eq:cv}
\ee
Consequently, $C(T)$ exhibits intricate behavior as a function of temperature and distinguishes the three dynamical regimes. More precisely, in the ultra-low temperature regime where $\alpha=1,$ Eq. (\ref{eq:cv}) implies that the specific heat is constant. However, as we enter the low temperature regime characterized by $\alpha<1$, $C(T)$ starts increasing as we increase temperature. Contrary to this, in the high temperature regime with $\alpha>1$, the specific heat intriguingly decreases with increasing temperature. Thus, this anomalous behavior of $C(T)$ clearly separates the different temperature regimes. Notably, this kind of anomalous behavior of specific heat has been observed in different context, from experimental measurements in certain liquid crystal films \cite{Jin_1996, Chou_1997, Chou_1998} and corresponding theoretical modeling using coupled hexatic-nematic XY model \cite{Touchette_2020}.

\begin{figure}[h]
 \centering \includegraphics[width=8 cm]{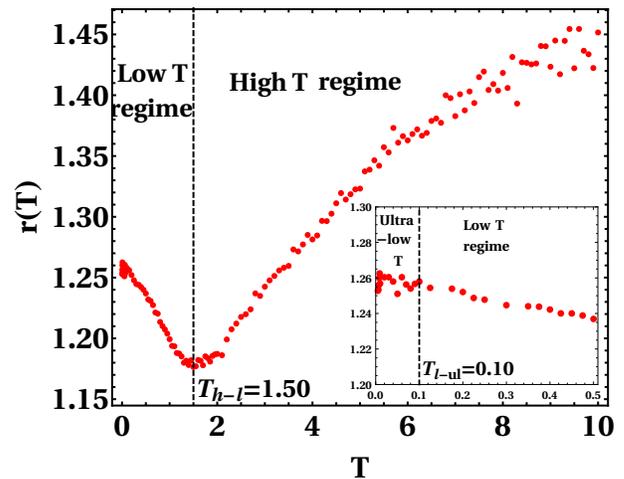} 
 \caption{The behavior of $r(T)$ (Eq. \ref{eq:rT}) evidently distinguishes the three dynamical regimes of DNLS. $r(T)$ increases monotonically with temperature in the high temperature regime, in contrary to its monotonically decreasing trend in the low temperature regime. The temperature at which minimum of $r(T)$ occurs, defines the crossover temperature $T_{h-l}$  between high temperature and low temperature regime. As shown in the inset, $r(T)$ remains almost constant in the ultra-low temperature regime followed by a monotonic decrease in the low temperature regime, the crossover temperature being marked as $T_{l-ul}$. Parameters used are $g=2.0$, $\mu=2.0$ and the chain length is $N=128$.}
 \label{fig:rT}
\end{figure}


Another equivalent and more transparent demarcator between different dynamical regimes is the ratio between $T$ and $\epsilon$, 
\be
r(T)\equiv\frac{T}{\epsilon}. 
\label{eq:rT}
\ee
This is a simpler demarcator as the three regimes can be identified by looking at the plot $r(T)$ versus $T$ as shown in Fig.~\ref{fig:rT}. From Fig.~\ref{fig:rT}, we observe that $r(T)$ increases monotonically with increasing temperature in the high temperature regime ($T>T_{h-l}$). In sharp contrast, in the low temperature regime ($T_{l-ul}< T< T_{h-l}$), $r(T)$ decreases monotonically with increasing temperature. Clearly, as a function of $T$, $r(T)$ shows a minimum at the crossover between low temperature and high temperature regime marked by $T_{h-l}$ (which is $1.50$ for the parameter values used in Fig.~\ref{fig:rT}). The distinction between low temperature and ultra-low temperature regime is shown in the inset of Fig.~\ref{fig:rT}. There, we observe that $r(T)$ is almost a constant with fluctuations of very small amplitudes in the ultra-low temperature regime ($0<T<T_{l-ul}$). However, once the temperature $T_{l-ul}$ (which is $0.10$ for the parameter values used in Fig.~\ref{fig:rT}) is reached, we start observing a monotonic decrease in $r(T)$ that marks the onset of low-temperature regime. 
It is thereby obvious that $r(T)$ is a much more transparent demarcator than $\alpha$ (Fig.~\ref{fig:difft}). In other words, $r(T)$  is a more effective and transparent way to find out the location of crossover temperatures as discussed below. 


\begin{table*}[ht]
\begin{center}
    \begin{tabular}{|p{0.20\textwidth}|p{0.20\textwidth}|p{0.10\textwidth}|p{0.30\textwidth}|}
    \hline
    \hspace*{0.1cm}\textbf{Dynamical regime} & \hspace*{0.1cm}\textbf{Temperature range} & \hspace*{0.1cm}$\mathbf{\alpha}$ & \hspace*{0.1cm} $r(T)$ \\
      \hline
       Ultra-low temperature & $T < T_{l-ul}$ & $=1$ & constant \\ 
     \hline
     Low-temperature & $T_{l-ul}<T<T_{h-l}$ & $<1$ & decreasing with increasing $T$ \\ 
     \hline
    High-temperature & $T>T_{h-l}$ & $\alpha>1$  &  increasing with increasing $T$ \\ 
     \hline
    \end{tabular}
     \caption{The table summarizes the three different dynamical regimes of the one dimensional DNLS in equilibrium with corresponding temperature ranges. In this table, we state how the properties of $\alpha$ (Eq. (\ref{eq:tphi}), Fig. \ref{fig:difft}) and $r(T)$ (Eq. (\ref{eq:rT}), Fig. \ref{fig:rT}) clearly demarcates these three dynamical regimes.}
     \label{tab:table}
     \end{center}
\end{table*} 

It is remarkable that the crossover temperature $T_{h-l}$ can be obtained exactly from the minimum of $r(T)$ vs. $T$ plot i.e.
\be
T_{h-l}=\underset{T}{\text{arg~min}} ~r(T).
\label{eq:criteria}
\ee
This finding based on our numerical results is  consistent with the criterion for determining the crossover temperature in Ref.~\onlinecite{Mendl_2015} obtained from a very different approach. This criterion \cite{Mendl_2015} is based on the frequency of specific dynamical processes leading to an additional conservation law (apart from that of the total energy and total mass) at low temperatures. We discuss this in detail in section~\ref{sec:phaseslip}.

We conclude this section with a brief summary of the main findings which is presented in Table \ref{tab:table}. In Table \ref{tab:table}, we observe that the exponent $\alpha$ in the $T-\epsilon$ relation (Eq.~(\ref{eq:tphi}), Fig.~\ref{fig:difft}) acts as a clear demarcator of the three different dynamical regimes of the DNLS in equilibrium. More precisely, we find that $\alpha=1$, $\alpha<1$ and $\alpha>1$ for ultra-low, low and high temperature regimes respectively. Another remarkable demarcator of the three regimes is the ratio $r(T)$ (Eq.~(\ref{eq:rT}), Fig.~\ref{fig:rT}). As exhibited in Table \ref{tab:table}, $r(T)$ remains constant in the ultra-low temperature regime, whereas it decreases monotonically in the low temperature regime and increases monotonically in the high temperature regime. Also, the crossover temperature $T_{h-l}$ is interestingly given by the temperature at which minimum of $r(T)$ occurs [Eq.~(\ref{eq:criteria})]. The prediction from this criterion has excellent agreement with the criterion proposed in Ref.~\onlinecite{Mendl_2015}, as will be discussed in detail in the next section.

\section{Emergence and disappearance of an almost conserved quantity: phase slips}
\label{sec:phaseslip}
In this section, we would like to probe the different dynamical regimes of DNLS through the emergence and disappearance of an additional almost conserved quantity, namely the total phase difference, as temperature is varied. To define this observable in a systematic way, let us consider the ground state ($T=0$) of the DNLS Hamiltonian in Eq.~(\ref{eq:hmu}), in terms of the original complex valued field $\psi_j^0$-s (where $0$ denotes the ground state), given by \cite{Iubini_2013}
\be
\psi_j^0(t)=\sqrt{\rho_0}e^{\mathrm{i}(\pi j + \mu t)}, ~~~j=1,2 \dots N.
\label{eq:psi0}
\ee
Here, $\rho_0$ is the average mass density calculated in ground state and $\mu$ is the chemical potential. 
At small non-zero temperature, the density field as well as the phase field will fluctuate over space and time. We denote them by $\rho_j(t)$ (background plus fluctuation) and $\theta_j(t)=\pi j+\mu t +\nu_j(t)$, respectively. Consequently, the $\psi$ fields also become non-trivially dependent on space and time as
\be 
\psi_j(t)=\sqrt{\rho_j}e^{\mathrm{i}(\pi j + \mu t+\nu_j)}, ~~~j=1,2 \dots N.
\label{eq:psij}
\ee
The variables $(\rho_j,\nu_j)$ are connected to the canonically conjugate variables $(q_j,p_j)$ introduced in Eq. (\ref{eq:qp}) as
\bea
q_j(t)&=&\sqrt{2\rho_j}~\mathrm{cos}(\pi j + \mu t+\nu_j) \cr
p_j(t)&=&\sqrt{2\rho_j}~\mathrm{sin}(\pi j + \mu t+\nu_j).
\label{eq:qjpj}
\eea
Consequently, in terms of the radial ($\rho_j$) and angle variables ($\nu_j$), the Hamiltonian in Eq. (\ref{eq:hmu}) can be expressed as
\be
H_\mu=\sum_{j=1}^{N}[-2 \sqrt{\rho_j \rho_{j+1}}~\mathrm{cos}(\nu_{j+1}-\nu_j)
+ \frac{g}{2} \rho_j^2 -\mu \rho_j ].
\label{eq:hmult}
\ee
The DNLS Hamiltonian in Eq.~(\ref{eq:hmult}), as already mentioned, has two conserved quantities: total energy $E$ and total mass $A$. This leads to local conservation laws in terms of conserved fields $\epsilon_j$ and $\rho_j$ (Table \ref{tab:observable}). Interestingly, it turns out that there is an additional emergent almost conserved quantity associated to the phase difference \cite{Mendl_2015}
\be
\delta\nu_j(t)=\mathrm{mod}[\nu_{j+1}(t)-\nu_j(t), 2\pi].
\label{eq:ps}
\ee
From Eq.~(\ref{eq:ps}), we note that the domain of $\delta\nu_j(t)$ is $[-\pi,\pi]$. As long as $\delta\nu_j(t)$ remains within $(-\pi, \pi)$, the total phase difference $\sum_j \delta \nu_j$ is conserved by the dynamics. This happens at low temperatures where one observes 
super-diffusive scaling of the  correlation function thereby putting it in the KPZ universality class \cite{Mendl_2015, Kulkarni_2013}.  As temperature is increased, phase difference $\delta \nu_j$ starts reaching the boundaries of the box $[-\pi, \pi]$ (i.e. $\delta\nu_j(t)=\pm \pi$). 
\begin{figure}[t]
  \centering
  \subfigure[]{\includegraphics[scale=0.57]{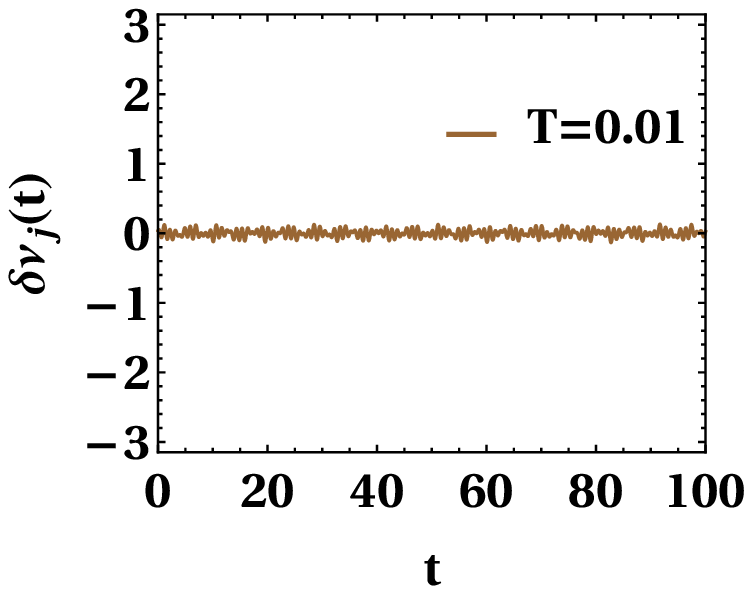}}\hfill
  \subfigure[]{\includegraphics[scale=0.57]{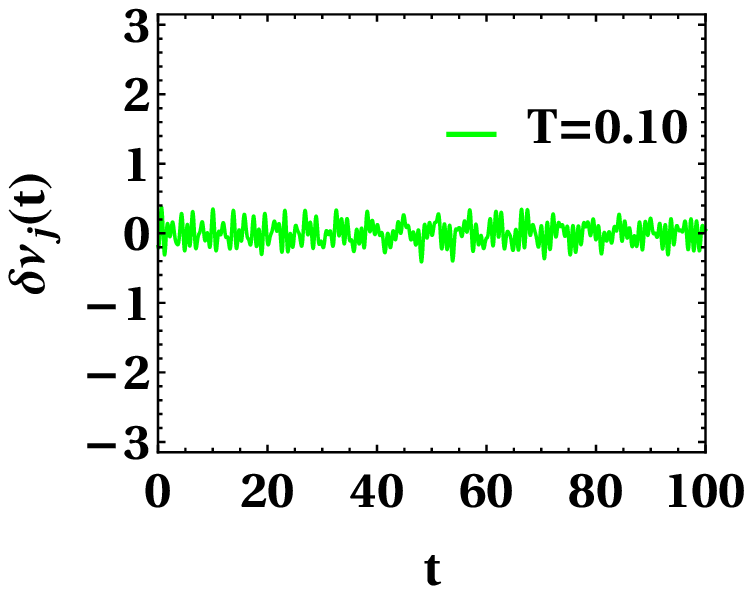}}\\
  \subfigure[]{\includegraphics[scale=0.57]{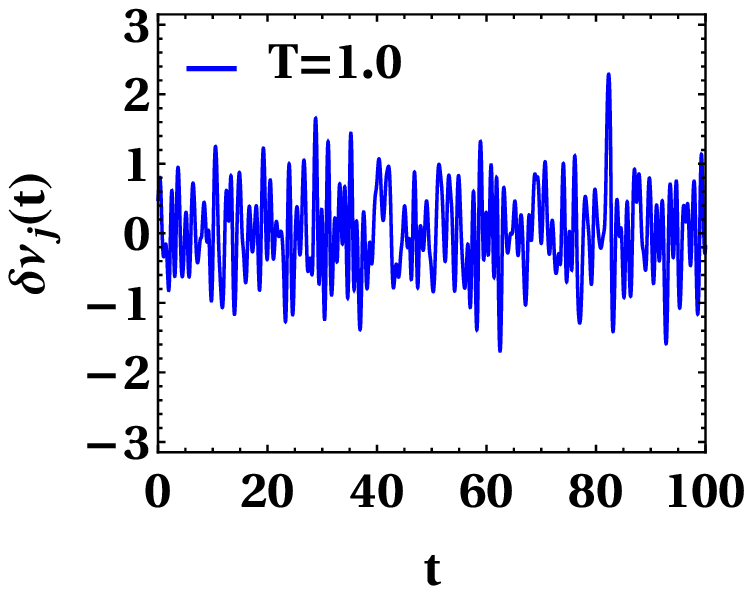}}\hfill
  \subfigure[]{\includegraphics[scale=0.57]{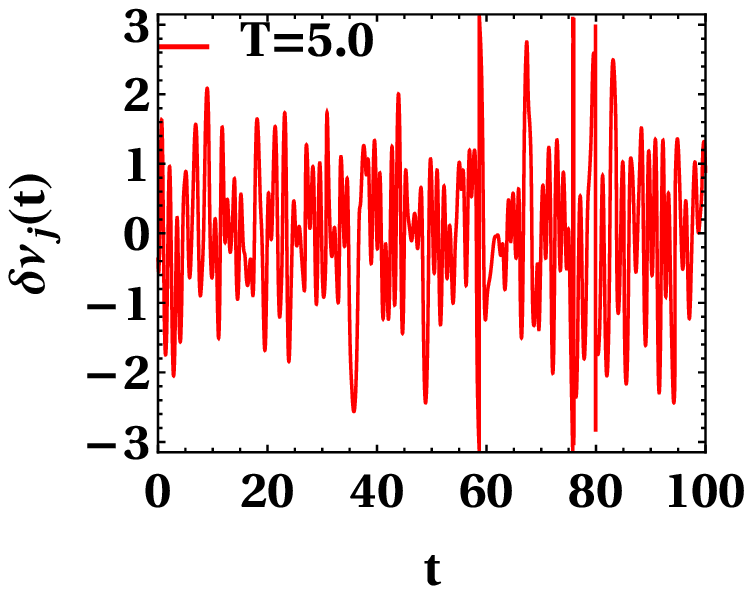}}
  \caption{The behavior of phase difference $\delta\nu_j(t)$ with time $t$ at few representative temperatures. Here $N=8$ and $j=4$ (i.e. the 4-th bond). At very low temperatures (a) $T$=0.01 and (b) $T$=0.10, the phase difference fluctuates very near to its ground state value $\delta\nu_j(t)=0$. At higher temperature (c) $T$=1.0, the phase difference increases considerably and at even higher temperature (d) $T$=5.0, $\delta\nu_j(t)$ touches the boundary values $\pm \pi$ at several times causing phase slips.}
\label{fig:psvst}
\end{figure}
This is exhibited in Fig.~\ref{fig:psvst} where we observe that at very low temperatures, the phase difference $\delta\nu_j(t)$ always remain very close to zero [Figs.~\ref{fig:psvst}(a), \ref{fig:psvst}(b)], thereby near the ground state value. However, as temperature is increased, $\delta \nu_j$ often takes bigger values [Fig.~\ref{fig:psvst}(c)] and at even higher temperatures [Fig.~\ref{fig:psvst}(d)], we notice that the phase difference starts touching the boundaries $\pm\pi$ at several times. At such an event, the winding number is increased ($\delta\nu_j(t)=-\pi$) or decreased ($\delta\nu_j(t)=+\pi$) by one unit. This results in discontinuous jumps of $\delta\nu_j(t)$ as a function of time. These jumps are called phase slips \cite{Mendl_2015,Das_2020}.
\begin{figure}[h]
 \centering \includegraphics[width=8 cm]{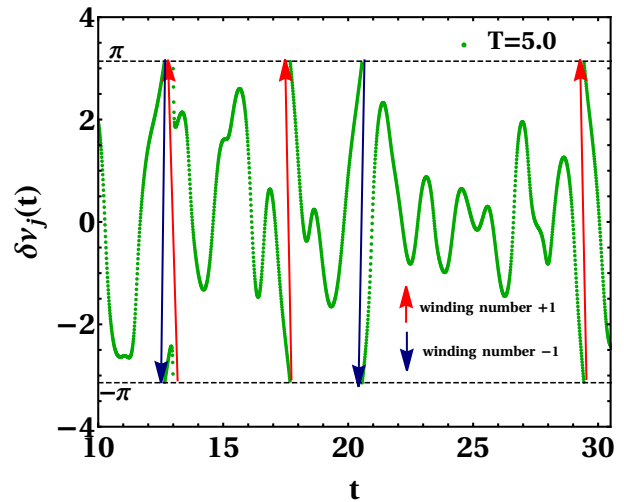} 
 \caption{The figure shows the occurrence of phase slip events for $j=4$ in a DNLS with $N=8$. Discontinuous jumps of amount $2\pi$ indicates phase slip events. In total there are five such events in this figure. In particular, the red arrows or $\delta\nu_j(t)=-\pi$ correspond to increase in winding number by $+1$ (three such events in the figure). The blue arrows or $\delta\nu_j(t)=\pi$ correspond to decrease in winding number by $-1$ (two such events in the figure). Therefore, up to the times  considered here, the net nonzero winding number is $(3-2)=1$, thereby providing a route for the breakdown of the emergent conservation law.}
 \label{fig:phaseslip}
\end{figure}
To illustrate the mechanism of the phase slip events, we present  $\delta\nu_j(t)$ versus $t$ in Fig.~\ref{fig:phaseslip} at $T=5.0$ which is sufficiently high to observe a good number of phase slips even within a small time interval. In Fig.~\ref{fig:phaseslip}, we observe total five discontinuous jumps or slip events (by an amount $2\pi$).
Among them, three events (red arrow) correspond to increase of winding number by $+1$ (i.e. $\delta\nu_j(t)=-\pi$ and $\delta\nu_j(t+\Delta t)=\pi$, $\Delta t$ being the appropriate time gap between two successive measurements). The remaining two (blue arrow) indicate decrease of winding number by $-1$ (i.e. $\delta\nu_j(t)=\pi$ and $\delta\nu_j(t+\Delta t)=-\pi$). Hence, during the short time interval in Fig.~\ref{fig:phaseslip}, the net nonzero winding number $(3-2=1)$ contributes to the breakdown of this emergent conservation law of the total phase difference through the bond $j$. Thus, phase slip events break the conservation of total phase difference.
\begin{figure}[t]
  \centering
  \subfigure[T=1.0]{\includegraphics[scale=0.37]{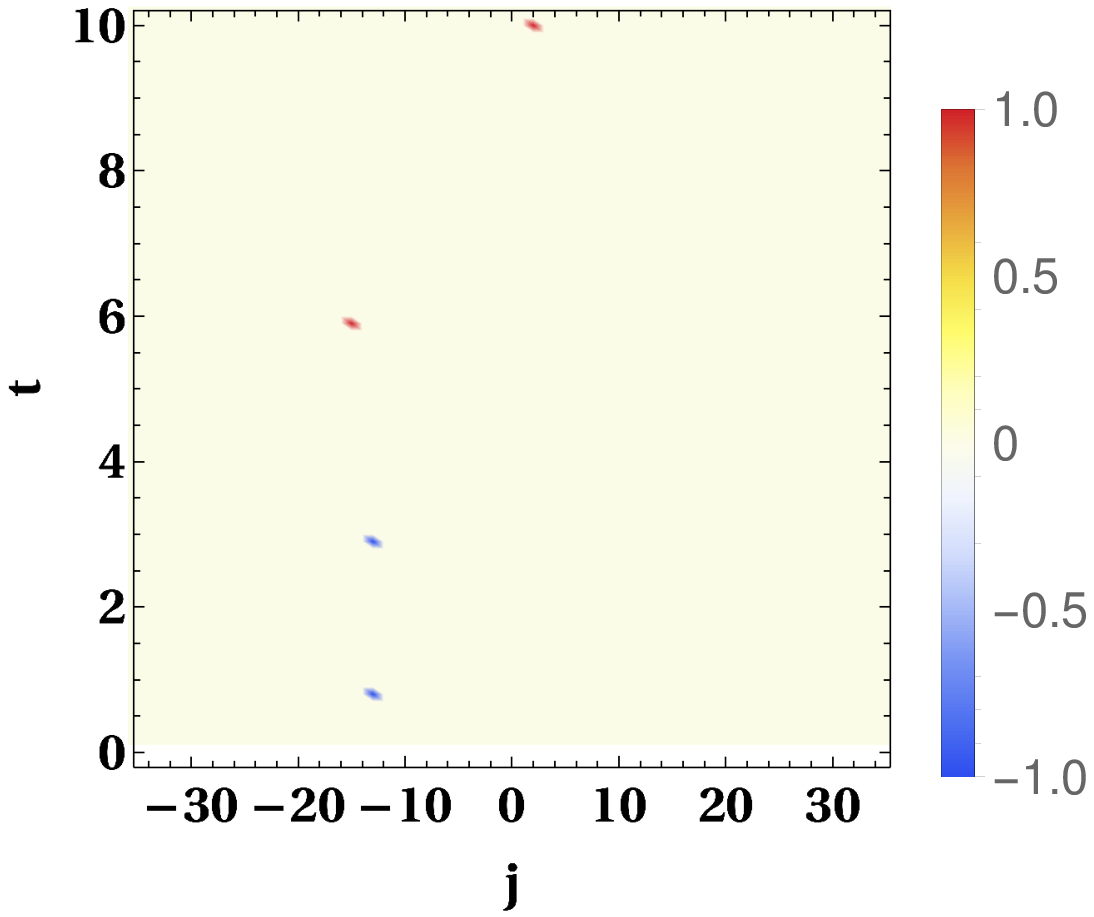}}\hfill
  \subfigure[T=1.5]{\includegraphics[scale=0.37]{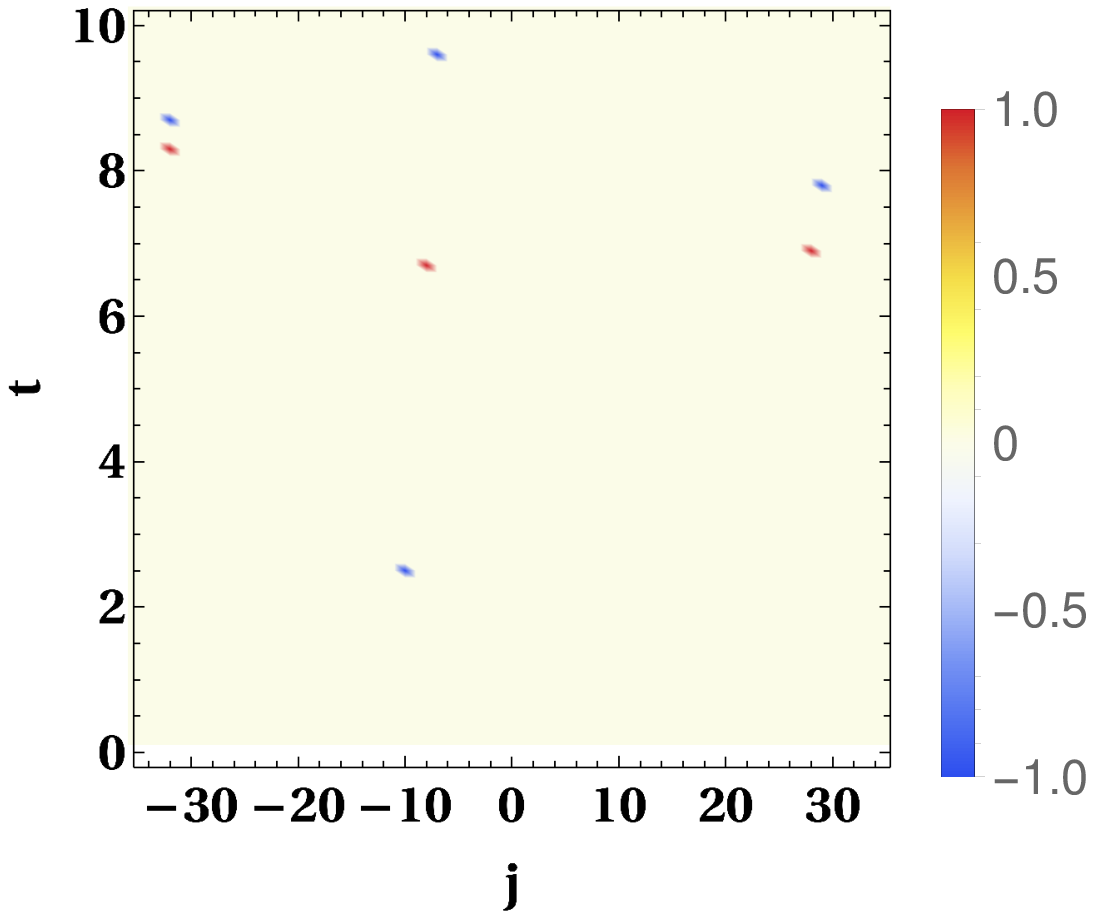}}\\
  \subfigure[T=2.0]{\includegraphics[scale=0.37]{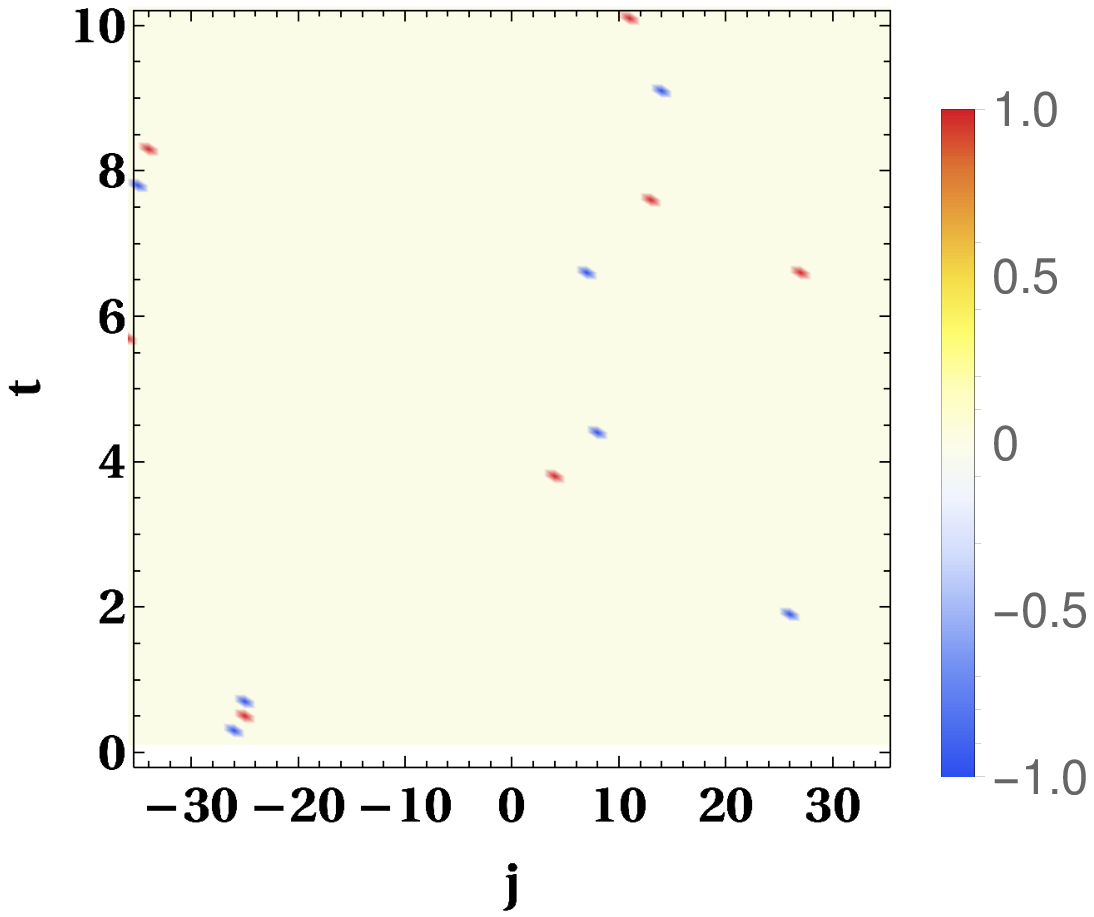}}\hfill
  \subfigure[T=4.0]{\includegraphics[scale=0.37]{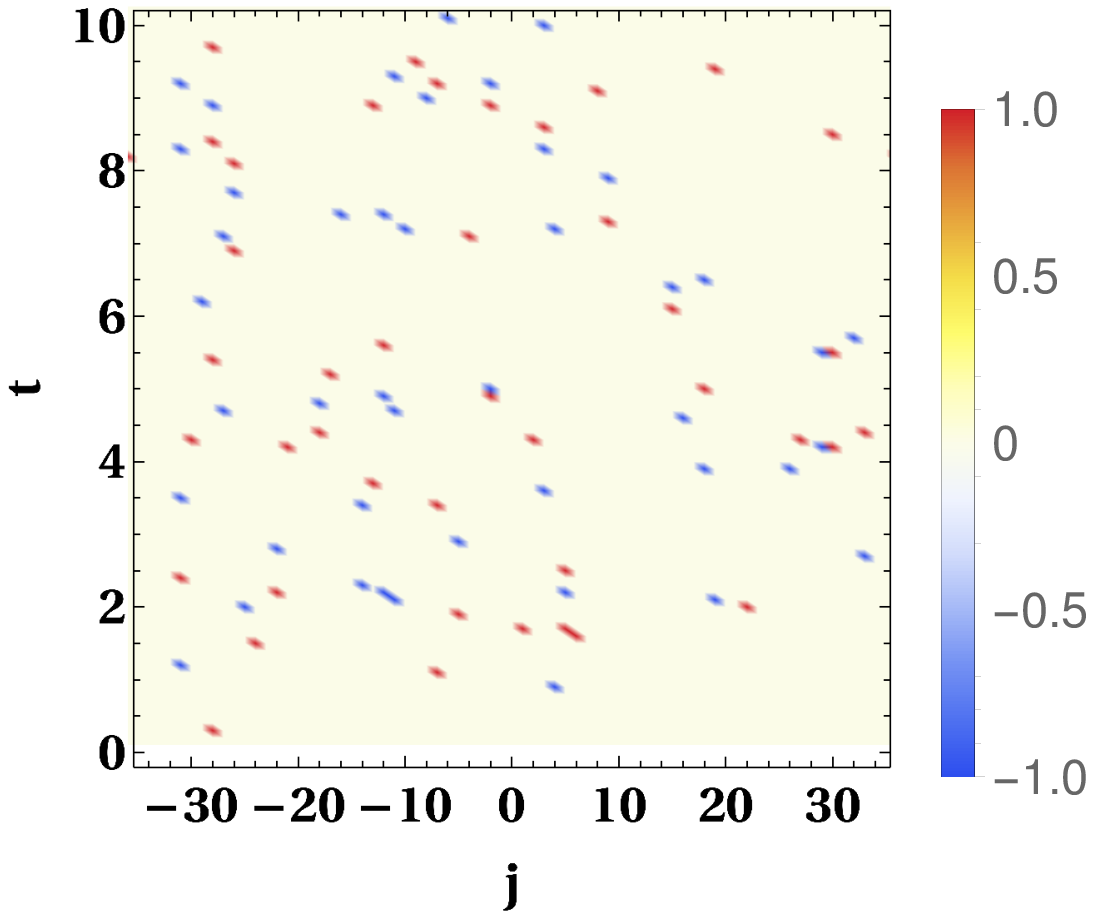}}
  \caption{The figure represents space-time heat-maps for phase slip events at different temperatures, measured in terms of change in winding numbers. The red and blue dots correspond to winding number +1 (i.e. $\delta \nu_j(t)=-\pi$) and -1 (i.e. $\delta \nu_j(t)=\pi$) respectively. While the phase slip events are very small in number at lower temperatures [(a) $T=1.0$, (b) $T=1.5$], they increase rapidly for comparatively higher temperatures [(a) $T=2.0$, (b) $T=4.0$]. The behavior of phase slips as a function of temperature is further analyzed in Fig.~\ref{fig:avrj} and Fig.~\ref{fig:pshtlt}. Here $N=128$, $\mu=2$  and $g=2$.}
\label{fig:rj}
\end{figure}
To observe how frequently these phase slip events happen in space$(j)$-time$(t)$ as we vary temperature, the spatio-temporal heat-maps for the winding numbers are presented in Fig.~\ref{fig:rj}. The red dots and blue dots in Fig.~\ref{fig:rj} correspond to unit increase (i.e. $\delta \nu_j(t)=-\pi$) and unit decrease (i.e. $\delta \nu_j(t)=\pi$) of winding numbers respectively. Fig.~\ref{fig:rj} clearly demonstrates that the total number of phase slip events are considerably small at low temperatures. As the temperature is increased, we observe noticeable increase in the number of phase slip events. Also, in the ultra-low temperature regime, we do not observe any phase slip events even for extremely long times considered here.

\begin{figure}[h]
 \centering \includegraphics[width=8 cm]{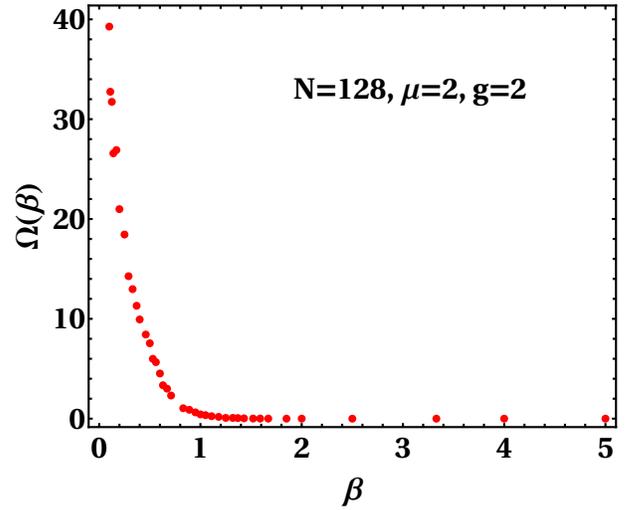} 
 \caption{Figure demonstrating that the average number of total phase slip events decreases exponentially with inverse temperature $(\beta)$. The system size is $N=128$ and the time up to which the phase slips (for all bonds) are counted is $t=500$. The number of samples over which the average is done is $20$.}
 \label{fig:avrj}
\end{figure}

Since these discontinuous jumps or phase slips occur randomly through dynamically activated processes, one can try to analyze them from the viewpoint of activation energy. 
If $\Delta V$ denotes the activation energy required on average to generate phase slips, the probability for such an event to occur is $\sim e^{-\beta \Delta V}$. This implies  that the  phase difference is conserved in the  low temperature regime up to a time scale $\sim e^{\beta \Delta V}$. Then the frequency of phase slip events is expected to vary with inverse temperature as $\Omega(\beta) \propto e^{-\beta \Delta V}$.
We numerically verify the exponentially decreasing nature of $\Omega(\beta)$ in Fig.~\ref{fig:avrj} where the total number of phase slip events on average is plotted against inverse temperature $\beta$. 
\begin{figure}[h]
 \centering \includegraphics[width=8 cm]{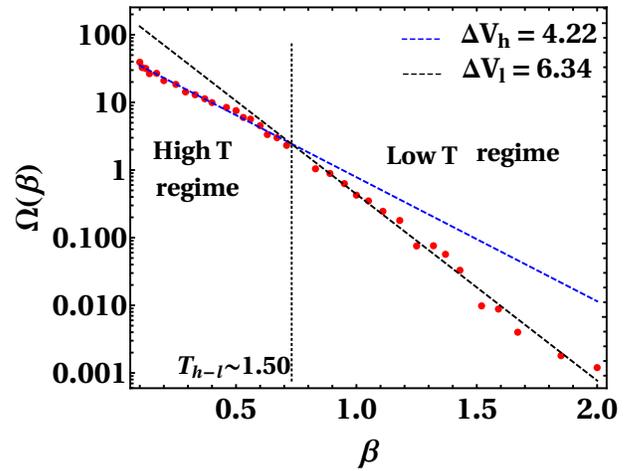} 
 \caption{Figure (log scale) exhibiting that the average activation energy in low temperature regime ($\Delta V_l=6.34$) is markedly higher than that of the high temperature regime ($\Delta V_h=4.22$). The crossover temperature between these two regimes is $T_{h-l}\sim1.50$ which is close to the $T_{h-l}$ value observed in Fig.~\ref{fig:rT}.}
 \label{fig:pshtlt}
\end{figure}
Further analysis of this exponential behavior is presented in Fig.~\ref{fig:pshtlt}. 
Interestingly, as shown in Fig.~\ref{fig:pshtlt}, the activation energy obtained from the slope of the $\Omega(\beta)$ versus $\beta$ plot in log scale, differs considerably in high temperature and low temperature regimes. We observe that the average activation energy for phase slips in low temperature regime $\Delta V_l=6.34$ is much higher than that of the high temperature regime $\Delta V_h=4.22$. 
The crossover temperature here $T_{h-l}\sim 1.50$ between two different regimes with different activation energies, is in good agreement with the crossover temperature $T_{h-l}=1.50$ observed from Fig.~\ref{fig:rT} and defined through Eq.~(\ref{eq:criteria}).

The numerically obtained $\Delta V_{h}$ and $\Delta V_{l}$ are results of complex dynamical processes along with averages. We would now like to get some theoretical insight into the activation energy. To do so, let us consider a phase slip event caused by the dynamics at a single site ($j$). Considering $\rho_j=\rho$ and $\delta \nu_j=\delta \nu$, the local energy at a single site is
\be
h(\rho,\delta \nu)=-2\rho~\mathrm{cos}(\delta \nu)+\frac{g}{2}\rho^2-\mu \rho.
\label{eq:hrhor}
\ee
It is straightforward to find that the minimum of $h(\rho,\delta \nu)$ in Eq. (\ref{eq:hrhor}) happens to be at $\delta \nu^*=0$ and $\rho^*=\left(\frac{\mu+2}{g}\right)$. Considering $\rho$ to be constant, the activation energy $\Delta V_1$ required for a phase slip event, i.e. $\delta \nu$ changing from $0$ to $\pm\pi$ , is
\be
\Delta V_1=h(\rho,\pm\pi)-h(\rho,0)=4\rho.
\label{eq:v1}
\ee
On the other hand, let us consider the case where $\delta \nu=\delta \nu^*=0$ (i.e. its ground state value). The notion of a phase slip at a given site becomes ill-defined when the density at that site goes to zero. The energy required to do so is given by
\be
\Delta V_2=h(0,0)-h(\rho^*,0)=\frac{g}{2}\rho^2.
\label{eq:v2}
\ee
Hence, we have estimates for $\Delta V_1$ [Eq.~(\ref{eq:v1})] and $\Delta V_2$ [Eq.~(\ref{eq:v2})]. These processes occur with frequencies $e^{-\beta \Delta V_1}$ and $e^{-\beta \Delta V_2}$ respectively. This in turn implies that the conservation of the phase difference in low temperature regime has lifetimes proportional to $e^{\beta \Delta V_1}$ and $e^{\beta \Delta V_2}$. 

To ensure that the total phase difference remains conserved for sufficiently long times, a safe estimate of $\beta \Delta V \gtrsim 2$ has been put forward in Ref.~\onlinecite{Das_2015} and Ref.~\onlinecite{Mendl_2015}. Here $\Delta V$ represents various mechanisms involved in phase slips. For example, in our case (DNLS), $\Delta V$ symbolizes $\Delta V_1$  [Eq.~(\ref{eq:v1})] and $\Delta V_2$  [Eq.~(\ref{eq:v2})].  In other words, the low temperature regime with three conservation laws are expected to prevail if both the following conditions are satisfied,
\be
\beta \Delta V_1 \gtrsim 2 \hspace*{1 cm} \& \hspace*{1 cm} \beta \Delta V_2 \gtrsim 2.
\label{eq:spohn1}
\ee
Using $\Delta V_1$ from Eq. (\ref{eq:v1}) and $\Delta V_1$ from Eq. (\ref{eq:v2}) (along with $g=2$), Eq.~(\ref{eq:spohn1}) becomes 
\be
\frac{2\rho}{T}>1 \hspace*{1 cm} \& \hspace*{1 cm} \frac{\rho^2}{2T} > 1.
\label{eq:spohn}
\ee
\begin{figure}[h]
\centering \includegraphics[width=8 cm]{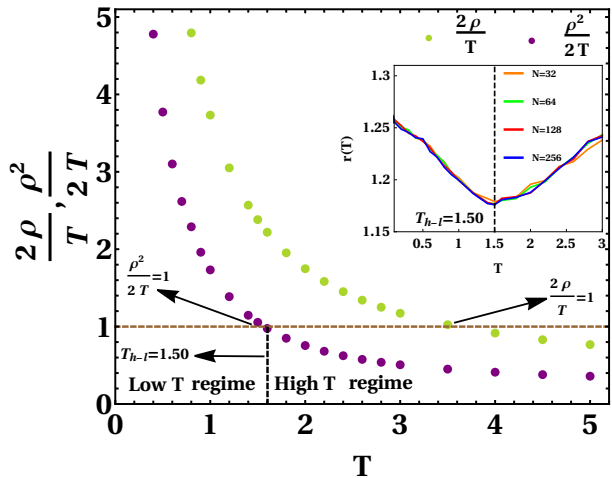}
\caption{The figure shows the range of temperature ($T<1.50$) in which the criterion Eq.~(\ref{eq:spohn}) holds, implying the DNLS is in the low temperature regime. We observe that the temperature at which either of the inequalities in Eq.~(\ref{eq:spohn}) begins to violate, marks the crossover temperature $T_{h-l}=1.50$ between low and high temperature regimes. To compare Eq.~(\ref{eq:spohn}) with the criterion we propose in Eq.~(\ref{eq:criteria}), we plot $r(T)$ with temperature in the inset, for different system sizes. The inset shows that the minimum of $r(T)$ occurs at the same temperature $T_{h-l}=1.50$, implying excellent agreement between the predictions from Eq.~(\ref{eq:spohn}) and 
Eq.~(\ref{eq:criteria}). Also, the inset of the figure exhibits satisfactory convergence of the $T_{h-l}$ value with increasing system size.}
\label{fig:compare}
\end{figure}
Clearly, the temperature at which the DNLS starts disobeying at least one of these inequalities, marks the crossover from low temperature to high temperature regime. To investigate Eq.~(\ref{eq:spohn}) numerically, we demonstrate the behavior of $\frac{2\rho}{T}$ and $\frac{\rho^2}{2T}$ as a function of temperature in Fig.~\ref{fig:compare}. We observe that both of the inequalities in Eq.~(\ref{eq:spohn}) are satisfied until one reaches $T_{h-l}\sim1.50$, implying $T<T_{h-l}$ to be the low temperature regime. At the crossover temperature, one of the conditions in Eq.~(\ref{eq:spohn}) begins to violate indicating the onset of high temperature regime $(T>T_{h-l})$. We note that the value of the crossover temperature $T_{h-l}$ estimated here is in excellent agreement to the corresponding $T_{h-l}$ value obtained from  Eq.~(\ref{eq:criteria}) which is presented in the inset of Fig.~\ref{fig:compare}. 

\begin{figure}[h]
 \centering \includegraphics[width=8 cm]{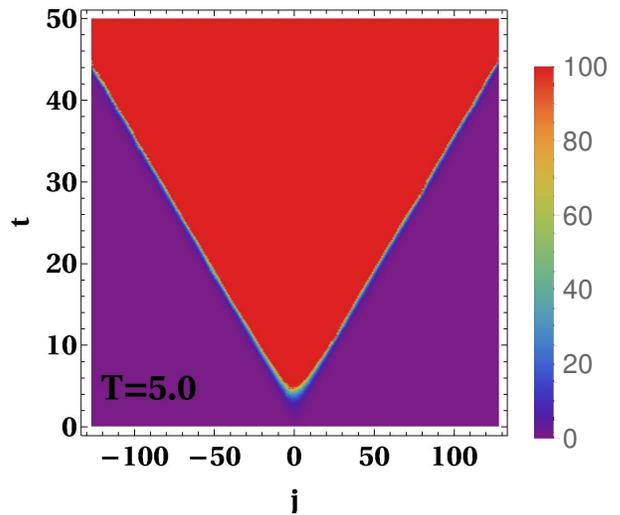} 
 \caption{Heat-map demonstrating ballistic spread of the OTOC (Eq.~\ref{eq:otoc}) creating a light-cone like structure where the sites inside the cone have  exponentially growing deviation. Here $N=256, \mu=2$ and $g=2$.}
 \label{fig:otoc}
\end{figure}

\section{OTOC, Lyapunov exponent and Butterfly speed in different regimes}
\label{sec:otoc}
In section \ref{sec:virial} we have discussed the distinction between different dynamical regimes of the DNLS based on one point macroscopic thermodynamic observables $T$, $\epsilon$ and their relation in Eq.~(\ref{eq:tphi}) (see Table \ref{tab:table}). Whereas in section \ref{sec:phaseslip}, these dynamical regimes are differentiated through the emergence and disappearance of an additional almost conserved quantity (total phase difference) caused by dynamically activated phase slip events. In this section, we would like to probe these regimes with a separate approach, based on observables related to many body chaos. It is worth recollecting that DNLS exhibits chaotic nature at high temperatures. At ultra low temperatures, DNLS is known to display almost integrable features \cite{Iubini_2012}.
Keeping in mind, the well-known connection between non-integrability and chaos \cite{Prigogine_1991,Masoliver_2011}, 
it would be interesting to investigate chaos in DNLS in different temperature regimes. To proceed along this direction, we investigate the classical out-of-time-ordered correlator (OTOC), the butterfly speed and the Lyapunov exponent \cite{Das_2018,Bilitewski_2018,Kumar_2019,Chatterjee_2020,Ruidas_2020,Bhanu_2020,Bilitewski_2020}. 

\begin{figure}[t]
 \centering \includegraphics[width=8 cm]{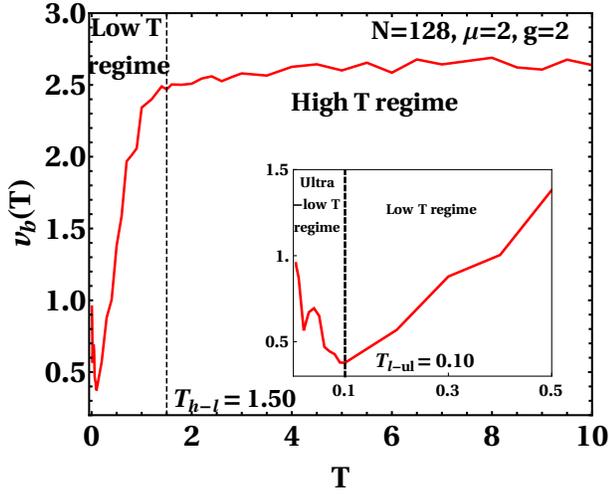} 
 \caption{Figure exhibiting distinctive behavior of the butterfly speed in the different temperature regimes. The main plot shows sharply increasing $v_b(T)$ with increasing temperature in the low temperature regime. This is separated from the very slowly varying speed in the high temperature regime, where the crossover temperature is around $T_{h-l}=1.50$. In the inset, we show the non-monotonic behavior of $v_b(T)$ between ultra-low temperature regime (the butterfly speed has an overall decreasing trend) and low temperature regime (sharply increasing speed). The occurrence of the minimum of the butterfly speed defines the crossover temperature $T_{l-ul}=0.10$ between these two regimes. This is consistent with the $T_{l-ul}$ value obtained previously in the inset of Fig.~\ref{fig:rT} ($r(T)$ versus $T$).}
 \label{fig:vbT}
\end{figure}

The OTOC is a spatio-temporal measure of chaos and in fact, both the butterfly  speed (spatial indicator of chaos) and the Lyapunov exponent (temporal indicator of chaos) can be derived directly from the OTOC \cite{Chatterjee_2020}. We define the OTOC for the microscopic degrees of freedom $q_j$-s (equivalently one can use $p_j$-s) as 
\be
 D_{q}(j,t;T) =\left\langle \left|\frac{q^I_j(t)-q^{II}_j(t)}{ q^I_0(0)-q^{II}_0(0)} \right| \right\rangle_{ic,T}= \left\langle \left|\frac{\delta q_j(t)}{\delta q_0(0)} \right| \right\rangle_{ic,T}.
\label{eq:otoc}
\ee
Here $\langle. \rangle_{ic,T}$ denotes average over initial conditions ($ic$) in equilibrium at temperature $T$ (see Appendix \ref{app:numerics}).
For notational convenience, we use $j=-N/2 +1, \dots,0 \dots,N/2$ in this section.
In Eq. (\ref{eq:otoc}), we consider two copies ($I$ and $II$) of the DNLS, which initially ($t=0$)  differ from each other only by an infinitesimal deviation $\delta q_0(0).$ We measure how this initially localized (at $0$-th site) deviation affects the system at other points in space at later time $t$. The explicit expressions for the equations of motion for $\delta q_j$-s and the details of numerical integration can be found in Appendix \ref{app:numerics}. Clearly, the OTOC $D_q(j,t;T)$ in Eq.~(\ref{eq:otoc}) is a function of space and time and we present the corresponding heat-map in Fig.~\ref{fig:otoc} at $T=5.0$ (high temperature regime). In Fig.~\ref{fig:otoc}, we observe a light-cone structure with sharp boundaries where the sites inside the light-cone have exponential growth of the deviation. 
This light-cone like OTOC-s are typically observed in other chaotic Hamiltonian systems \cite{Das_2018, Bilitewski_2018,Bilitewski_2020}. 

\begin{figure}[t]
 \centering \includegraphics[width=8 cm]{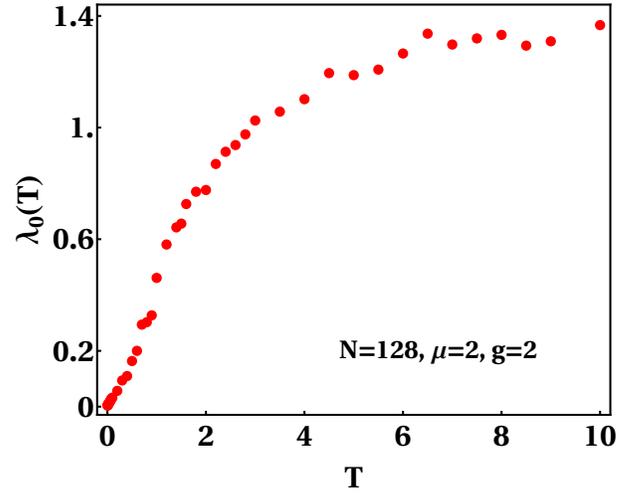} 
 \caption{Figure exhibiting the monotonically increasing characteristic of the Lyapunov exponent with increasing temperature. Interestingly, $\lambda_0(T)$ increases much faster with $T$ at low temperatures, compared to its slower growth rate at high temperatures. Further analysis of this behavior is presented in Fig.~\ref{fig:fitlT}.}
 \label{fig:lT}
\end{figure}

\begin{figure*}
  \centering
  \subfigure[]{\includegraphics[scale=0.38]{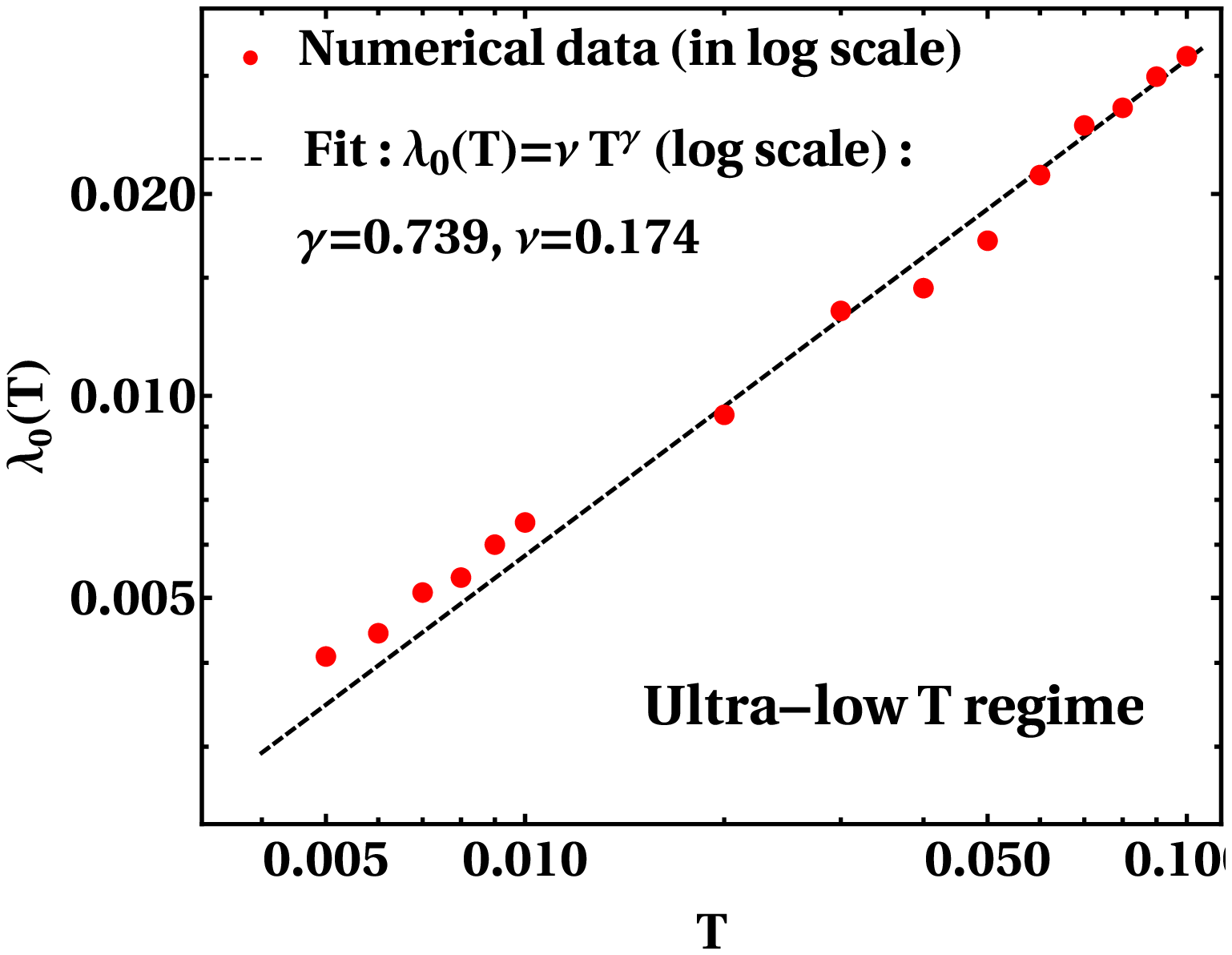}}\hfill
  \subfigure[]{\includegraphics[scale=0.38]{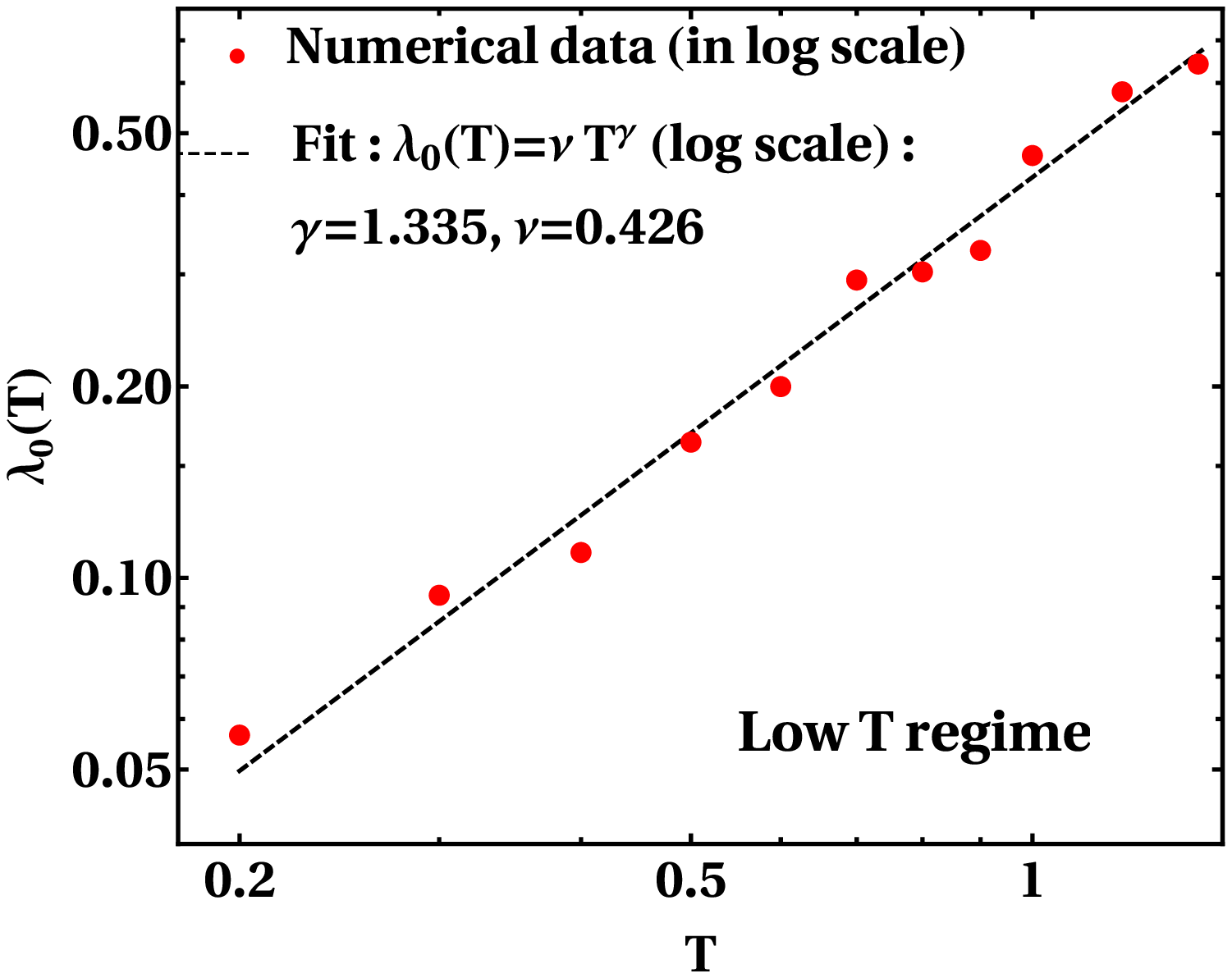}}\hfill
  \subfigure[]{\includegraphics[scale=0.38]{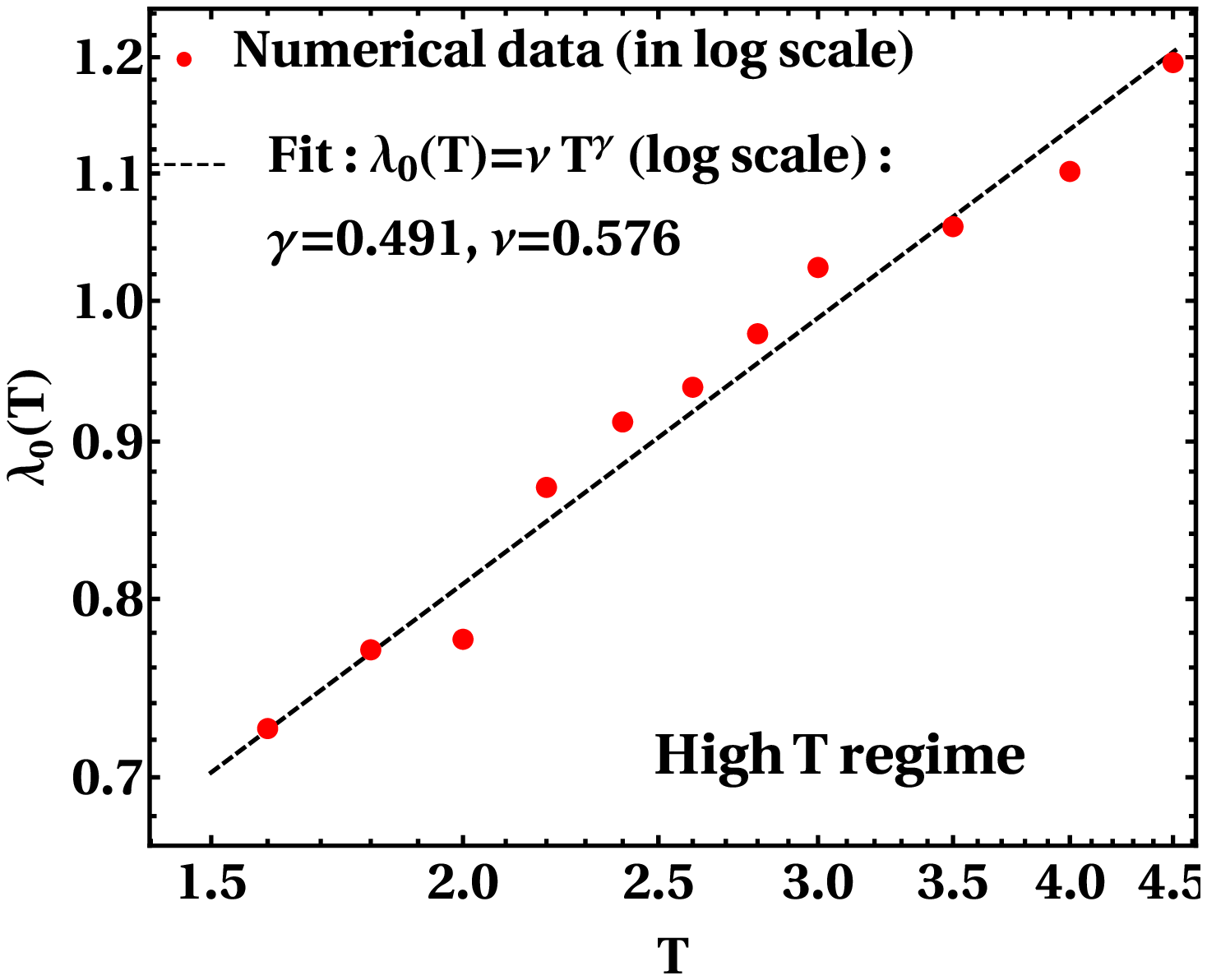}}\\
  \caption{In this figure, the numerical data for Lyapunov exponent versus temperature is fitted to a power law of the form $\lambda_0(T)=\nu T^\gamma$. Figures (a), (b) and (c) show significant variation in the value of the exponent $\gamma$ along the crossovers between different dynamical regimes. Notably, the growth rate of $\lambda_0(T)$ as a function of $T$ is largest in the low temperature regime as seen in (b). The behavior $\lambda_0(T) \sim \sqrt{T}$ in the high temperature regime [as seen in (c)], has been observed previously in different contexts \cite{Bilitewski_2018,Ruidas_2020,Kumar_2019,Kumar_2020}.}
\label{fig:fitlT}
\end{figure*} 

The ballistic spread of the light-cone in Fig.~\ref{fig:otoc} implies the existence of a constant speed of spatial propagation of the OTOC, known as the butterfly speed. As mentioned earlier, the butterfly speed can be defined directly using the OTOC, given below
\be
v_b(T)=\frac{1}{t}\left\langle\sum_{j=1}^{N}\Theta\left(\frac{\delta q_j(t)}{\delta q_0(0)}-1\right)\right\rangle_{ic,T}.
\label{eq:vb}
\ee
The step function $\Theta(.)$ in Eq.~(\ref{eq:vb}) measures how many sites have gained deviation greater than or equal to the initial deviation (at site $0$) after some time $t$. From Fig.~\ref{fig:otoc} we observe that this number grows proportional to $t$. Therefore, Eq.~(\ref{eq:vb}) gives us the constant speed $v_b(T)$ which depends on the temperature $T$. The behavior of the butterfly speed as we vary the temperature, is shown in Fig.~\ref{fig:vbT}. Interestingly, we observe that the butterfly speed behaves very differently in the three distinct dynamical regimes. 
As shown in the inset of Fig.~\ref{fig:vbT}, $v_b(T)$ has an overall decreasing trend with increasing $T$ in the ultra-low temperature regime. In sharp contrast, the butterfly speed increases rapidly as $T$ increases in the low temperature regime as observed in Fig.~\ref{fig:vbT} and its inset.
This non-monotonic behavior of $v_b(T)$ with $T$, defines the crossover temperature $T_{l-ul}$ between ultra-low and low temperature regimes. More precisely, $T_{l-ul}$ can be measured as the temperature at which minimum of the butterfly speed occurs, given by
\be
T_{l-ul}=\underset{T}{\text{arg~min}} ~v_b(T).
\label{eq:criteria-lt-ult}
\ee
Notably, the value of $T_{l-ul}$ obtained in the inset of Fig.~\ref{fig:vbT} is in very good agreement with the same observed in the inset of Fig.~\ref{fig:rT} where a different observable $r(T)$ [Eq. (\ref{eq:rT})] has been investigated. 
A non-monotonic characteristic of $v_b(T)$, similar to the one observed here, has been reported recently \cite{Ruidas_2020} in the context of classical 2D XXZ model. There,  the minimum of the butterfly speed occurs at the transition temperatures for both the Ising and  the Kosterlitz-Thouless transitions. 
The steepness of the growth of the butterfly speed as a function of temperature, falls off considerably as soon as the system enters to the high temperature regime. This is observed from the very slowly varying trend of $v_b(T)$ in the high temperature regime in Fig.~\ref{fig:vbT}. The crossover temperature between the slowly varying butterfly speed in high temperature regime and the rapidly increasing speed in low temperature regime, happens to be around $T_{h-l}=1.50$. This agrees very well with the crossover temperature $T_{h-l}$ in Fig.~\ref{fig:rT} ($r(T)$ versus $T$) and Fig.~\ref{fig:pshtlt} ($\Omega(\beta)$ versus $\beta$), obtained previously using very different approaches.

\begin{figure}[h]
  \centering
  \subfigure[T=0.005]{\includegraphics[scale=0.39]{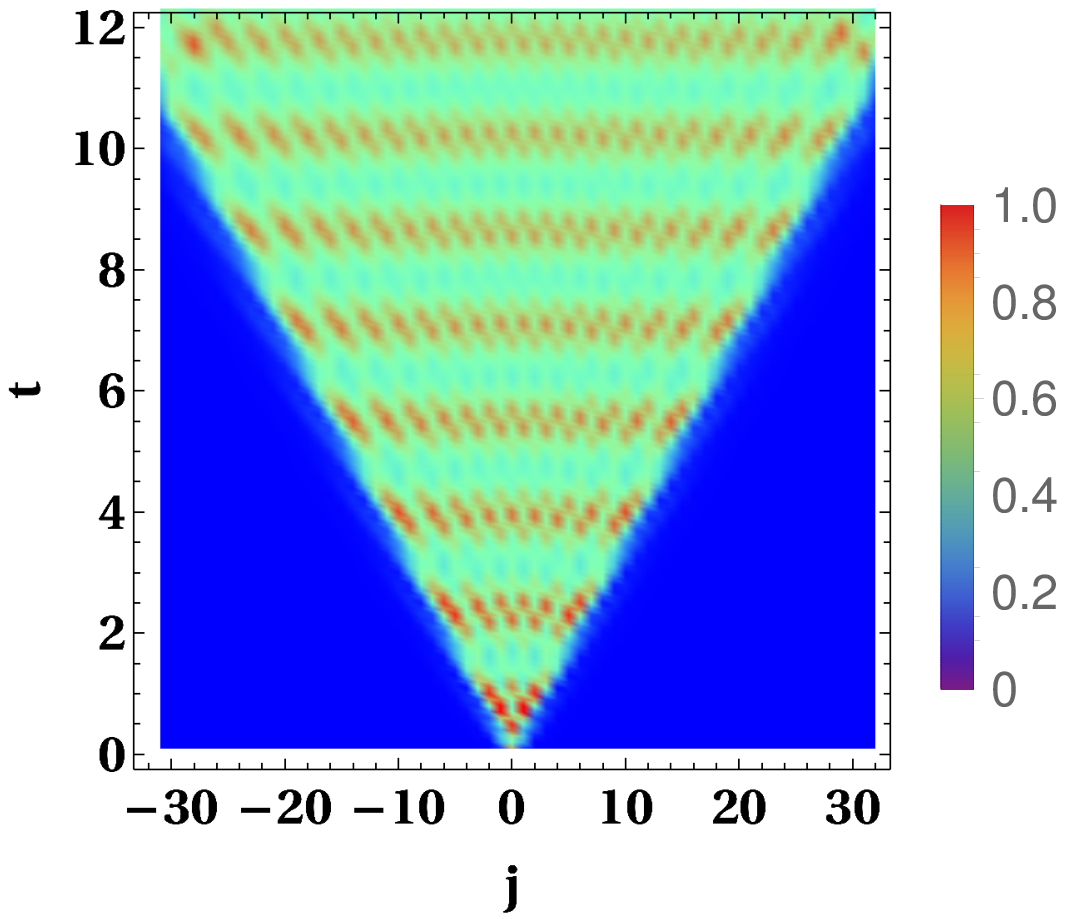}}\hfill
  \subfigure[T=1.0]{\includegraphics[scale=0.39]{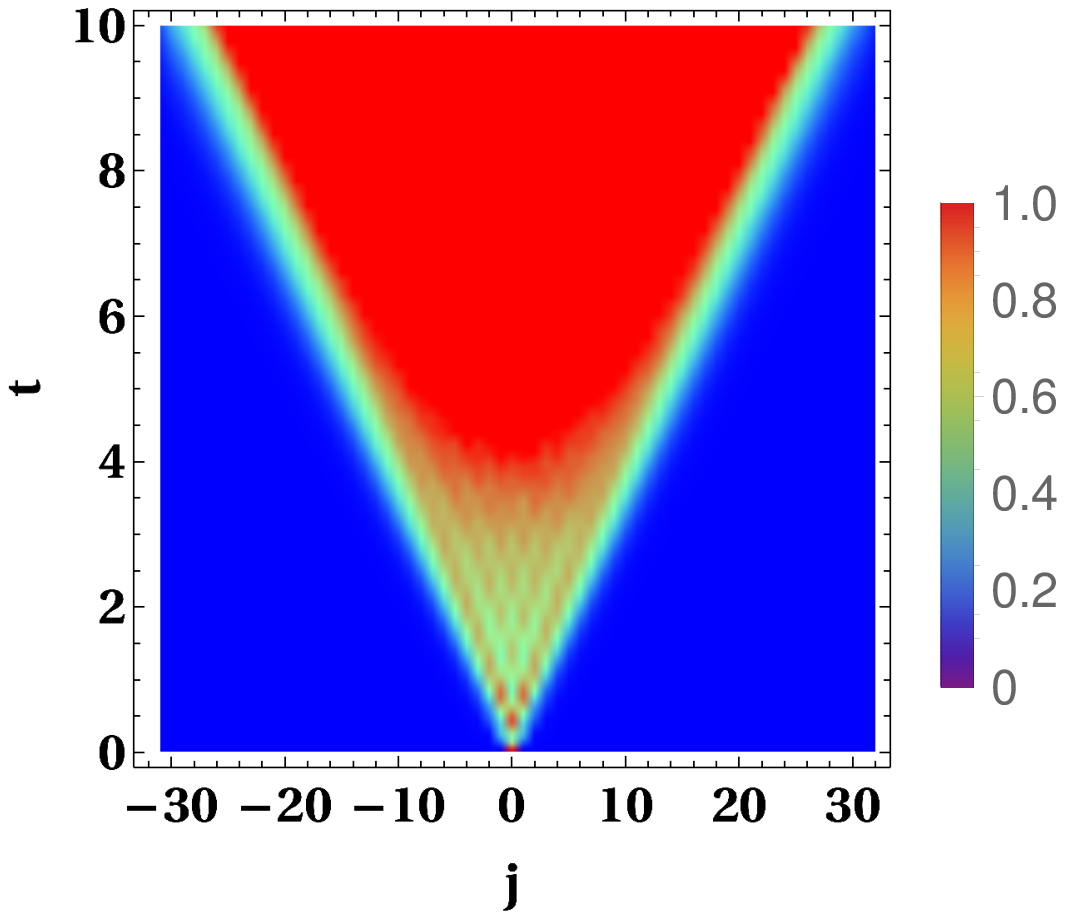}}
  \caption{Figure demonstrating striking differences between the space($i$)-time($\tau$) heat-maps of OTOC [Eq.~(\ref{eq:otoc})] at (a) $T=0.005$ (ultra-low temperature regime) and (b) $T=1.0$ (low temperature regime). In the ultra-low temperature regime (a), we observe intriguing oscillatory structures both in time and space. This feature vanishes in the low temperature regime regime (b). Also, (b) shows exponential growth of OTOC inside the light-cone. The typical OTOC behavior in the high temperature regime, exhibiting exponential growth inside the light-cone has already been presented in Fig.~\ref{fig:otoc} $(T=5.0)$.}
\label{fig:otochm}
\end{figure}

To analyze the temporal growth of the OTOC (Fig.~\ref{fig:otoc}) at temperature $T$, we consider the Lyapunov exponent [$\lambda_j(T)$] at the site $j$  defined as
\be
\lambda_j(T)=\left\langle \mathrm{lim}_{t\rightarrow \infty} \frac{1}{t}~\mathrm{ln}\left|\frac{\delta q_j(t)}{\delta q_0(0)}\right|\right\rangle_{ic,T}.
\label{eq:lambda}
\ee
At sufficiently large time, each $\lambda_j(T)$ (where $j=-N/2 +1, \dots,0 \dots,N/2$) converges to the same constant value \cite{Chatterjee_2020}. Therefore, without any loss of generality, one can focus on the behavior of $\lambda_0(T)$ as temperature is varied. This is presented in Fig.~\ref{fig:lT}. The figure shows that $\lambda_0(T)$ is a monotonically increasing function of $T$. However, a more careful observation reveals that the Lyapunov exponent grows much faster with $T$ at lower temperatures in comparison to a slower growth at sufficiently high temperatures. 
To investigate this behavior in a more systematic way, we fit the numerically obtained $\lambda_0(T)$ to the power law $\nu\,T^\gamma$ separately in the three dynamical regimes. The corresponding results are plotted on log scales in Fig.~\ref{fig:fitlT}. 
We observe that the value of the exponent $\gamma$ deviate significantly in the different dynamical regimes. As expected from the observations in Fig.~\ref{fig:lT}, we find that $\gamma$ is maximum in the low temperature regime [Fig.~\ref{fig:fitlT}(b)] indicating the maximum growth rate of the Lyapunov exponent in this regime. Interestingly, as demonstrated in Fig.~\ref{fig:fitlT}(c), $\gamma \sim 0.5$ in the high temperature regime. We should mention that this behavior $\lambda_0(T)\sim\sqrt{T}$ has also been observed recently in some other chaotic Hamiltonian systems with very different microscopic dynamics \cite{Bilitewski_2018,Ruidas_2020,Kumar_2019,Kumar_2020}. 

Having discussed how the butterfly speed and the Lyapunov exponents display significantly different behaviors in different dynamical regimes, 
a natural question that arises is, how the heat-maps of OTOC in these regimes differ from each other. 
To answer this, we present in Fig.~\ref{fig:otochm} the corresponding OTOC heat-maps in ultra-low and low temperature regimes (recall that Fig.~\ref{fig:otoc} represents the typical OTOC behavior in high temperature regime). Indeed, from Fig.~\ref{fig:otochm}(a) we observe interesting  oscillatory structures in both space and time, thereby manifesting the almost integrable nature of DNLS in the ultra-low temperature regime. This is in sharp contrast to the behaviors in low temperature [Fig.~\ref{fig:otochm}(b)] and high temperature regimes [Fig.~\ref{fig:otoc}].

\section{Summary}
\label{summary}
In this paper, we have shown that the one dimensional discrete nonlinear Schr\"{o}dinger chain, which has an interesting non-separable Hamiltonian structure, exhibits three different dynamical regimes at finite temperatures. These three regimes, namely the ultra-low, low  and high temperature regimes, have been characterized here and differentiated from one another using several distinct approaches. These include (i) analyzing one point macroscopic thermodynamic observables (temperature $T$, energy density $\epsilon$) and their relationship ($T\propto\epsilon^\alpha$), (ii) investigating the emergence and disappearance of an additional (apart from total energy and total mass) almost conserved quantity (total phase difference) by studying phase slip events, and (iii) probing the chaotic dynamics of the DNLS with the classical out-of-time-ordered correlators and derived quantities (butterfly speed, Lyapunov exponent).

The nontrivial task of thermalizing the non-separable DNLS Hamiltonian [Eq.~(\ref{eq:hamqp})] has been achieved here by connecting the system to Langevin thermostats at both ends following the procedure [Eq.~(\ref{eq:eqm})] in Ref.~\onlinecite{Iubini_2013}. Through rigorous numerical simulations we verify that this nontrivial process indeed leads to proper thermalization in the DNLS (Fig.~\ref{fig:vir_can}). We find numerically that the one point thermodynamic observables, namely temperature ($T$) and energy density ($\epsilon$), defined in Table \ref{tab:observable}, follow the relation $T=c\,\epsilon^\alpha$. Remarkably, $\alpha$ acts as a prominent identifier of the three different dynamical regimes. 
More precisely, we notice that $\alpha=1$ in the ultra-low temperature regime, $\alpha<1$ in the low temperature regime and $\alpha>1$ in the high temperature regime (Fig.~\ref{fig:difft}). The demarcation of these different regimes becomes even more visibly clear from the behavior of $r(T)=T/\epsilon$ as a function of temperature (Fig.~\ref{fig:rT}). To elaborate, $r(T)$ remains almost constant in the ultra-low temperature regime, decreases monotonically in the low temperature regime and increases monotonically in the high temperature regime. 
This overall non-monotonic behavior of $r(T)$ helps us to identify the crossover temperature $T_{h-l}$ (between high and low temperature regimes) which remarkably turns out to be the temperature at which minimum  of $r(T)$ occurs [Eq.~(\ref{eq:criteria})]. The characterization of the three different regimes using $\alpha$ and $r(T)$ has been summarized in Table \ref{tab:table}.

The DNLS has two conserved quantities, the total energy and the total mass [Eq.~(\ref{eq:hamqp})]. Interestingly, an additional almost conserved quantity, namely the total phase difference emerges in the low temperature regime making it distinct from the high temperature regime (where this conservation does not hold). This emergence and disappearance of the third conservation law has been analyzed here through the concept of dynamical processes leading to discontinuous jumps or phase slip events (Fig.~\ref{fig:phaseslip}). In fact, the total number of phase slip events falls off exponentially as a function of the inverse temperature (Fig.~\ref{fig:avrj}). Importantly, we find that the activation energy required for the phase slip events are significantly different in the low and high temperature regimes, thereby demarcating these two regimes (Fig.~\ref{fig:pshtlt}). 
This in turn means that the frequency of phase slip events are very low in the low temperature regime, resulting in the conservation of the total phase difference for extremely long times. On the other hand, the phase slips occur very frequently in the high temperature regime (Fig.~\ref{fig:rj}). We also find the in the ultra-low temperature regime, phase slips do not occur even at very long times that we have considered here. 
The theoretical estimates of the activation energies involved in the phase slip events predicts the crossover temperature $T_{h-l}$ which is found to be in excellent agreement (Fig.~\ref{fig:compare}) to that obtained from the previous approach (Eq.~(\ref{eq:criteria}), Fig.~\ref{fig:rT}).

To probe the chaotic nature of the DNLS at high temperature and its almost integrable behavior in the ultra-low temperature regime, we investigate the classical OTOC [Eq.~(\ref{eq:otoc})], butterfly speed [Eq.~(\ref{eq:vb})] and Lyapunov exponent [Eq.~(\ref{eq:lambda})]. In particular, the butterfly speed exhibits an interesting non-monotonic behavior with varying temperature (Fig.~\ref{fig:vbT}). It has an overall decreasing behavior in the ultra-low temperature regime contrary to the rapidly increasing characteristic in the low temperature regime, followed by a much slower growth rate in the high temperature regime. Remarkably, the crossover temperature $T_{l-ul}$ can be measured as the temperature at which the minimum of the butterfly speed occurs (Eq.~(\ref{eq:criteria-lt-ult}), Fig.~\ref{fig:vbT}). On the other hand, the Lyapunov exponent increases monotonically as a function of temperature (Fig.~\ref{fig:lT}). However, it shows interesting crossovers in the values of the exponent $\gamma$ when fitted to a power law $\lambda_0(T)=\nu T^\gamma$ (Fig.~\ref{fig:fitlT}). Particularly, the maximum growth rate of the Lyapunov exponent with temperature happens to be in the low temperature regime. This is followed by a behavior $\lambda_0(T)\sim\sqrt{T}$ in the high temperature regime that has been observed previously in other contexts \cite{Bilitewski_2018,Ruidas_2020,Kumar_2019,Kumar_2020}. The space-time heat-maps of the OTOC presents visibly prominent differences between the ultra-low temperature regime (Fig.~\ref{fig:otochm}(a), oscillatory structures in space-time inside the light-cone), low [Fig.~\ref{fig:otochm}(b)] and high temperature regimes (Fig.~\ref{fig:otoc}). 

Having established these various methods, it would be interesting to adapt them to explore other interacting many body systems including non-separable Hamiltonian systems (e.g. various generalizations of DNLS \cite{Sarma_2014,Ablowitz_2014,Ablowitz_2016,Mithun_2021}, spin chains \cite{Das_2018,Das_2020,Ishimori_1982,Roberts_1988,Nowak_2015} etc.). In future, we plan to understand the different dynamical regimes and the onset of chaos in such systems through the lens of a  mode coupling theory \cite{Bilitewski_2020}.

 \section*{Acknowledgements}
We thank Avijit Das for useful discussions. MK  would  like  to  acknowledge support from the project 6004-1 of the Indo-French Centre for the Promotion of Advanced Research (IFCPAR),
Ramanujan  Fellowship  (SB/S2/RJN-114/2016),  SERB Early Career Research Award (ECR/2018/002085) and SERB Matrics Grant (MTR/2019/001101) from the Science and Engineering Research Board (SERB), Department  of  Science  and  Technology,  Government  of  India.
AK  would  like  to  acknowledge the SERB Early  Career  Research  Award  ECR/2017/000634  from the Science and Engineering Research Board,  Department of Science and Technology, Government of India. The numerical calculations were done on the clusters {\it Mario} and  {\it Tetris} at the ICTS-TIFR. We acknowledge support of the Department of Atomic Energy, Government of India, under Project No.RTI4001.
 
\appendix
\section{Details of numerical procedure}
\label{app:numerics}
Here we provide the numerical details for (i) initial conditions, (ii) equations of motion, (iii) methods of numerical integration, and (iv) averaging procedures used in this paper to compute the observables of interest. To start with, we recall that the effect of the Langevin thermostats (used for thermalization)  has to be invoked carefully because of the non-separable nature of the DNLS Hamiltonian [Eq.~\ref{eq:hamqp}]. Unlike the separable Hamiltonians where the interaction with Langevin thermostats modifies only the momentum equations, equations for both $q$ and $p$ get modified for DNLS \cite{Iubini_2012}. In other words, $q$ and $p$ for DNLS are on equal footing. This makes the procedure for achieving equilibration using Langevin thermostats, significantly different from the traditional approach. The equations of motions take the following form,
\bea
\dot{q_1}&=&\frac{\partial H}{\partial p_1}-\gamma \frac{\partial H_\mu}{\partial q_1}+\sqrt{2\gamma T}\, \xi_{1}'(t)\cr&=&\bar{f}(q_1,p_1,q_2,p_2)+ \sqrt{2\gamma T}\, \xi_{1}'(t)\cr
 \dot{p_1}&=&-\frac{\partial H}{\partial q_1}-\gamma \frac{\partial H_\mu}{\partial p_1}+\sqrt{2\gamma T}\, \xi_{1}''(t)\cr&=&\tilde{f}(p_1,q_1,p_2,q_2)+\sqrt{2\gamma T}\, \xi_{1}''(t)\cr\cr
 \dot{q_j}&=&\frac{\partial H}{\partial p_j}=f(p_{j+1},p_{j-1},p_j,q_j),\;\;\;\; j=2 \dots (N-1)\cr
 \dot{p_j}&=&-\frac{\partial H}{\partial q_j}=-f(q_{j+1},q_{j-1},q_j,p_j),\;\;\; j=2 \dots (N-1) \cr\cr
 \dot{q_N}&=&\frac{\partial H}{\partial p_N}-\gamma \frac{\partial H_\mu}{\partial q_N}+\sqrt{2\gamma T}\, \xi_{N}'(t)\cr&=&\bar{f}(q_N,p_N,q_{N-1},p_{N-1})+ \sqrt{2\gamma T}\, \xi_{N}'(t) \cr
 \dot{p_N}&=&-\frac{\partial H}{\partial q_N}-\gamma \frac{\partial H_\mu}{\partial p_N}+\sqrt{2\gamma T}\, \xi_{N}''(t)\cr&=&\tilde{f}(p_N,q_N,p_{N-1},q_{N-1})+\sqrt{2\gamma T}\, \xi_{N}''(t),
 \label{eq:eqom}
\eea
where $\gamma$ is the coupling strength between the system and the bath. Since the Langevin thermostats are connected to both ends (i.e. $1^{\mathrm{st}}$ and $N^{\mathrm{th}}$ sites), the equations of motions for $q_1,p_1;q_N,p_N$ are modified accordingly in Eq.~(\ref{eq:eqom}). $\xi_{1}',\xi_{1}'',\xi_{N}',\xi_{N}''$ are Gaussian white noises each of which is delta correlated with unit variance i.e. $\langle\xi(t)\xi(s)\rangle=\delta(t-s)$. The explicit expressions for $\bar{f}(.),\tilde{f}(.),f(.)$ are listed below,
\bea
\bar{f}(x_1,x_2,x_3,x_4)&=&x_4+\frac{g}{2}x_2\left(x_2^2+x_1^2\right)\cr&-&\gamma\left[x_3+\frac{g}{2}x_1\left(x_2^2+x_1^2\right)-\mu x_1\right]\cr
\tilde{f}(x_1,x_2,x_3,x_4)&=&-x_4-\frac{g}{2}x_2\left(x_2^2+x_1^2\right)\cr&-&\gamma\left[x_3+\frac{g}{2}x_1\left(x_2^2+x_1^2\right)-\mu x_1\right]\cr
f(x_1,x_2,x_3,x_4)&=& x_1+x_2+\frac{g}{2}x_3\left(x_3^2+x_4^2\right).
\eea
We have used random initial conditions such that $(q_j,p_j)\in [-1,1]$ $\forall j$ at $t=0$. For numerical integration of the stochastic differential equations in Eq.~(\ref{eq:eqom}), we have utilized an improved version of the stochastic integration method (described in Ref.~\onlinecite{Mannella_2002}) based on Taylor series expansion that keeps terms of order $h^2$ for the deterministic terms and order $h^\frac{5}{2}$ for the stochastic terms. The time step size $h$, used for numerical integration, has been fixed to $0.001$.

For computing the OTOC [Eq.~\ref{eq:otoc}], butterfly speed [Eq.~\ref{eq:vb}] and Lyapunov exponent [Eq.~\ref{eq:lambda}] with average over initial conditions, we first let the system to thermalize at desired temperature $T$ using the procedure in Eq.~(\ref{eq:eqom}). Once the system reaches equilibrium, we detach the Langevin thermostats. Thereafter, we use fourth order Runge-Kutta method for numerically integrating the following equations of motion of the equilibrated DNLS system,
\bea
\dot{q_j}&=&f(p_{j+1},p_{j-1},p_j,q_j) \cr
 \dot{p_j}&=&-f(q_{j+1},q_{j-1},q_j,p_j) \cr \cr
\dot{\delta q_j}&=& g(\delta p_{j+1},\delta p_{j-1},\delta p_j,\delta q_j,p_j,q_j) \cr
\dot{\delta p_j}&=&-g(\delta q_{j+1},\delta q_{j-1},\delta q_j,\delta p_j,q_j,p_j), 
\label{eq:dqdp}
\eea
where $j=1,2,\dots,N$ and the explicit expressions for $f(.)$ and $g(.)$ are respectively,
\bea
f(x_1,x_2,x_3,x_4)&=& x_1+x_2+\frac{g}{2}x_3\left(x_3^2+x_4^2\right) \cr
g(x_1,x_2,x_3,x_4,x_5,x_6)&=& x_1+x_2+\frac{g}{2} x_3 (x_5^2+x_6^2)\cr&+&gx_4x_5x_6 .
\label{eq:g}
\eea 
We choose $\delta q_j(0)=\varsigma \delta_{j,k}$ and $\delta p_j(0)=0$, with $\varsigma=10^{-6}$ used in all the simulations. The average over the initial conditions has been done over $10^3$ equilibrated initial conditions.  

\begin{figure}[h]
  \centering
  \subfigure[]{\includegraphics[scale=0.30]{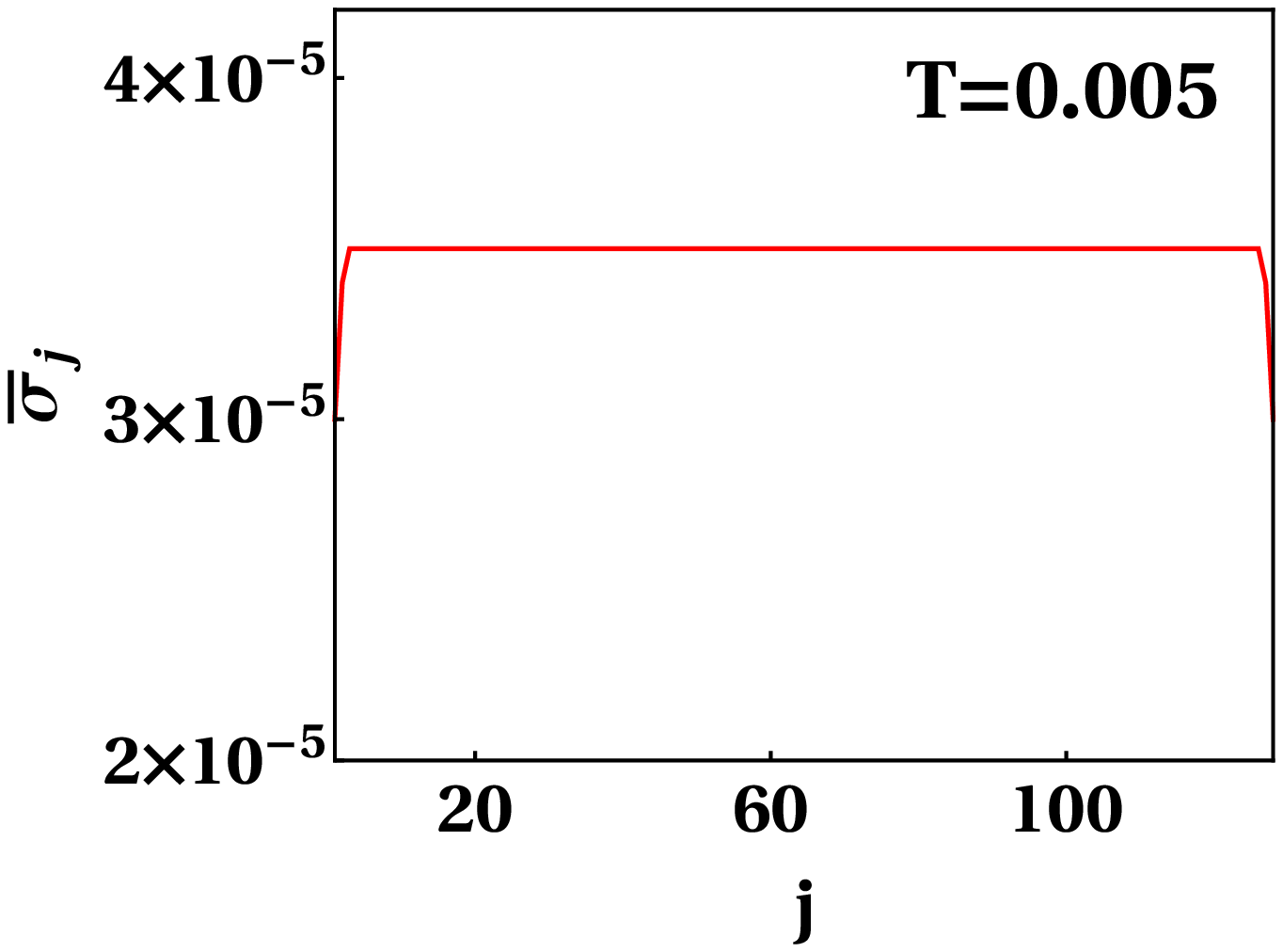}}\hfill
  \subfigure[]{\includegraphics[scale=0.30]{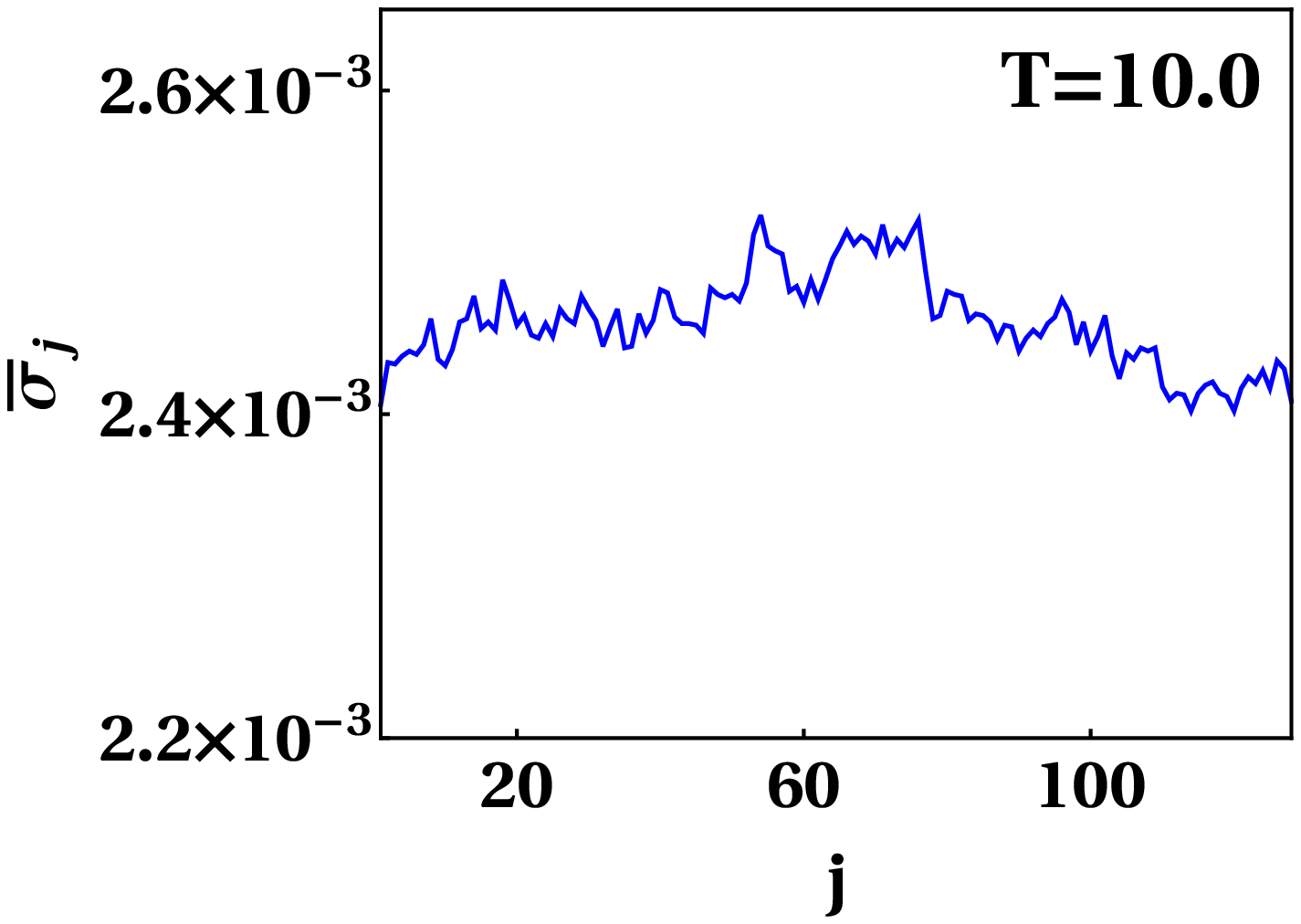}}
  \caption{Spatial profiles of the standard deviation $\bar{\sigma}_j$ [Eq.~(\ref{eq:sigma})] for the time averaged data set at  (a) $T=0.005$ (ultra-low temperature regime) and (b) $T=10.0$ (high temperature regime).}
\label{fig:sigmaj}
\end{figure}

\begin{figure}[t]
 \centering \includegraphics[width=8 cm]{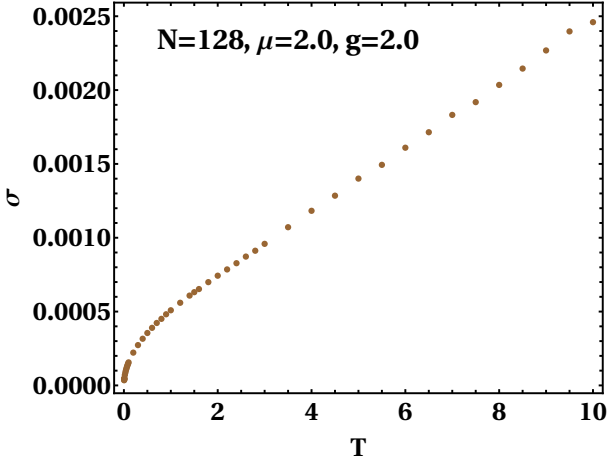} 
 \caption{The standard deviation $\sigma$ [Eq.~(\ref{eq:sigmabar})] obtained after doing both temporal and spatial average over the sample, plotted as a function of temperature.}
 \label{fig:sigmaT}
\end{figure}

\section{Computation of error bars for $\alpha$}
\label{app:error}
In this section, we would like to discuss in detail the error bars corresponding to the values of $\alpha$ [Eq.~(\ref{eq:tphi})] in different dynamical regimes (Fig.~\ref{fig:difft}). As shown in Fig.~\ref{fig:difft}, $\alpha=0.999(7)$ in ultra-low temperature regime, $\alpha=0.957(7)$ in low temperature regime and $\alpha=1.139(6)$ in high temperature regime ($\mu=2.0$, $g=2.0$). Since the differences between the values of $\alpha$ in different regimes are small, we would like to present here a careful and detailed analysis of the error bars associated with the corresponding $\alpha$ values. Below, we discuss this step by step.

\begin{enumerate}
\item The error bar $\mathrm{d}\alpha$ for the exponent $\alpha$ can be obtained by differentiating the relation $T=c\,\epsilon^\alpha$ [Eq.~(\ref{eq:tphi})] as
\be
\mathrm{d}\alpha=\left|\frac{\alpha\, \mathrm{d}\epsilon}{\epsilon\, \mathrm{ln}(\epsilon)}\right|,
\label{eq:dalpha}
\ee
where the temperature $T$ is kept fixed and $c$ is a constant. Note, in Eq.~(\ref{eq:dalpha}), we focus only on the absolute value of $\mathrm{d}\alpha$ since ultimately we would consider $\alpha\pm\mathrm{d}\alpha$. Clearly, to compute $\mathrm{d}\alpha$, we have to compute the standard deviation of the energy density $\mathrm{d}\epsilon$ from our simulations. 

\begin{figure}[t]
 \centering \includegraphics[width=8 cm]{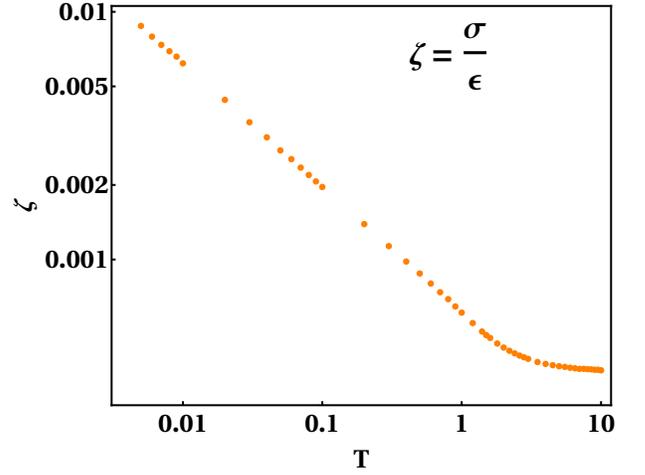} 
 \caption{Figure showing the relative error $\zeta=\sigma/\epsilon$  presented as a function of temperature in log-log scale. We observe $\zeta\ll 1$ in all the temperature regimes.} 
 \label{fig:nuT}
\end{figure}

\item  We start with a random initial condition for the microscopic variables $q_j$-s and $p_j$-s of the DNLS and let the system evolve. After the transient period is over, we start computing the observables of interest. In particular, here we consider the local energy density $\epsilon_j$ (Table \ref{tab:observable}) where $j=1,2 \dots N$. 
Adapting the ideas from Ref.~\onlinecite{Young_2014}, we do not compute the observable at each time step. Rather, we do that after every $100$ time steps to minimize the correlation between the data points of the sample.  

\item We collect a sample of total $n$ (here $n=1.5 \times 10^7$) data  points. The $m$-th data point in the sample would be denoted as $\epsilon_{j,m}$ where $m=1,2,\dots n$. Then we calculate the sample average $\bar{\epsilon}_j$ of energy density and the corresponding sample standard deviation $\bar{s}_j$ as
\bea
\bar{\epsilon}_j&=&\frac{1}{n}\sum_{m=1}^{n} \epsilon_{j,m} \cr
\bar{s}_j&=&\sqrt{\frac{1}{n}\sum_{m=1}^{n} (\epsilon_{j,m})^2-\langle\epsilon_{j,m}\rangle^2}.
\label{eq:s}
\eea
Note that $\bar{(.)}$ symbol denotes average over time-steps.

\begin{figure}[h]
  \centering
  \subfigure[]{\includegraphics[scale=0.33]{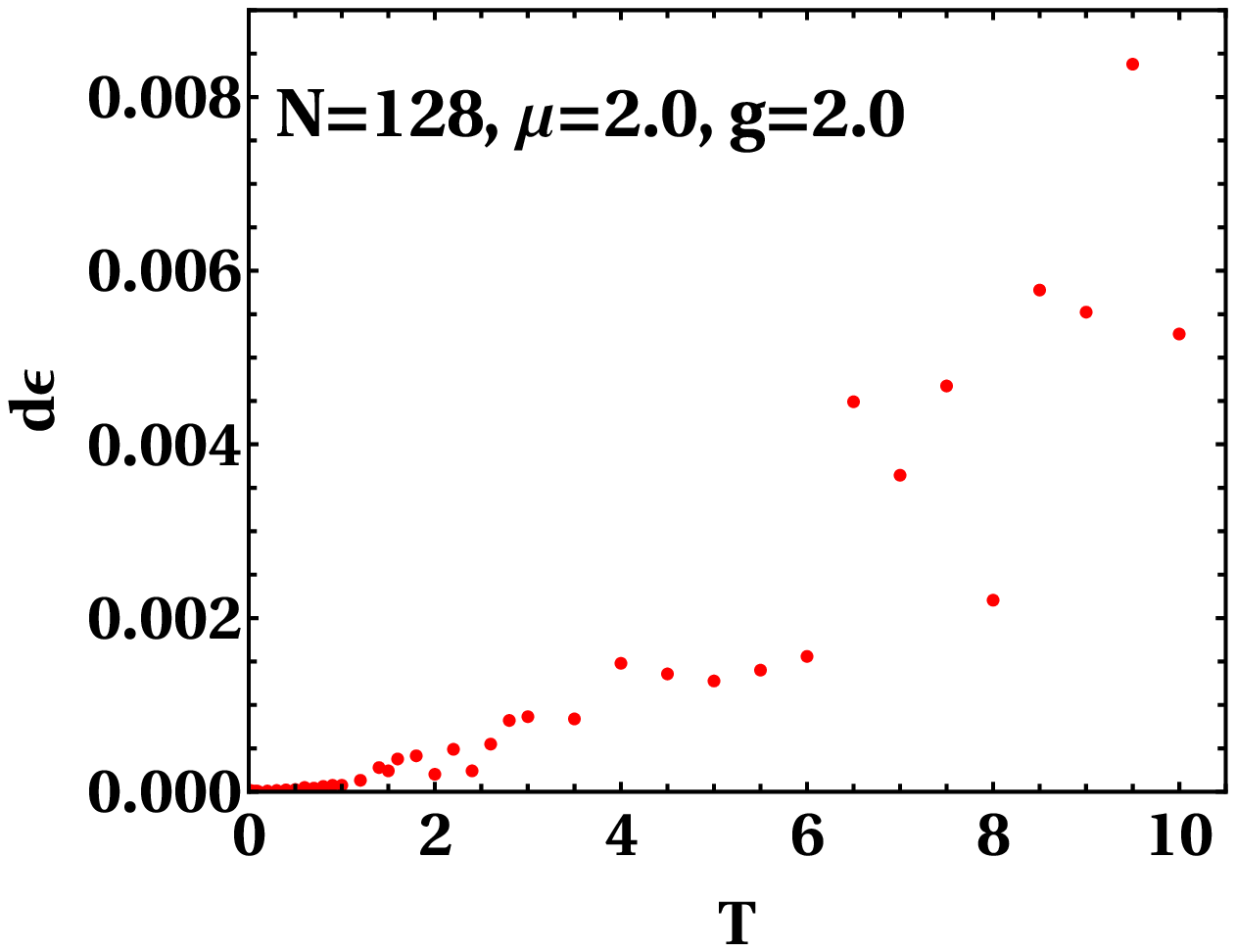}}\hfill
  \subfigure[]{\includegraphics[scale=0.33]{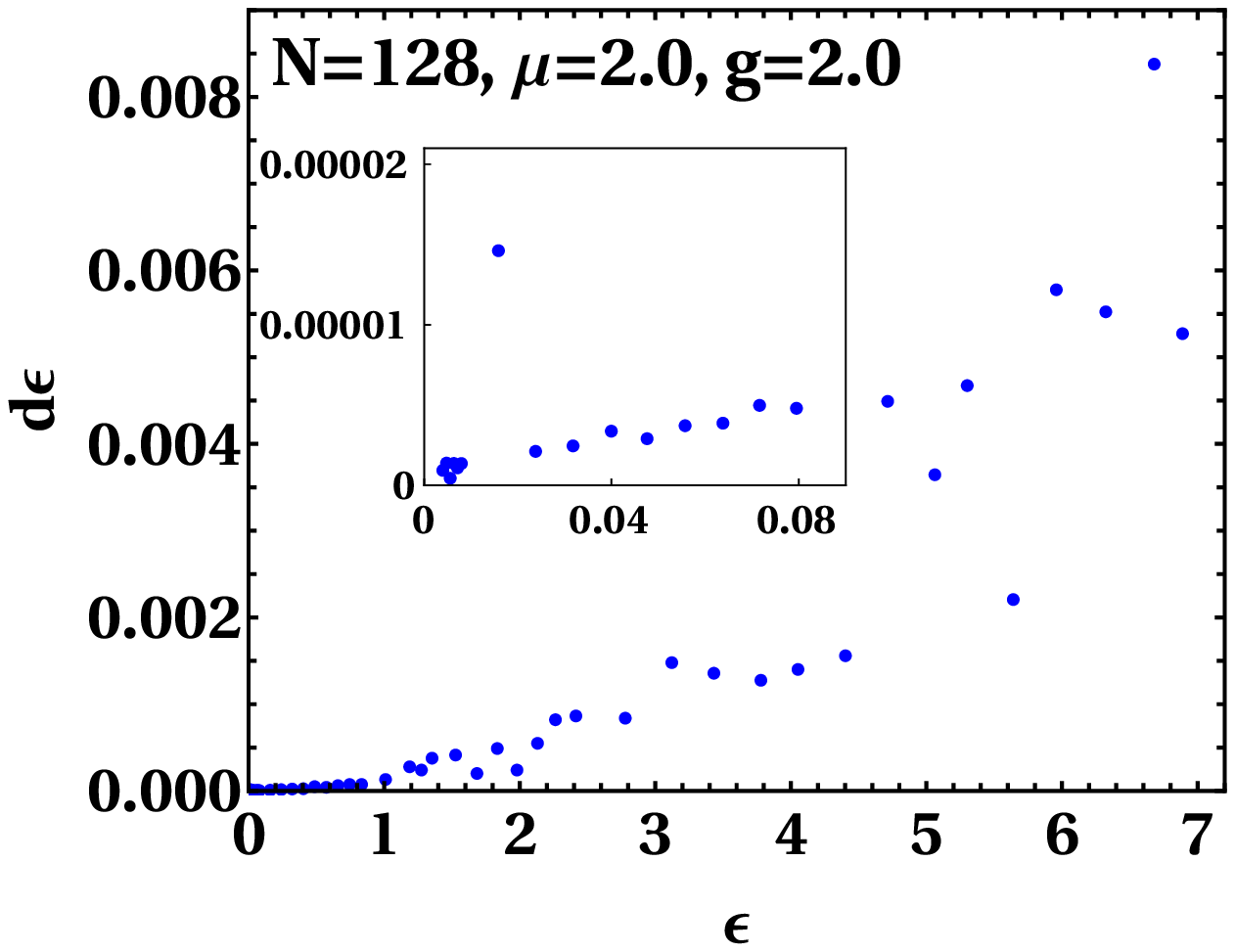}}
  \caption{Here we present $\mathrm{d}\epsilon$ [Eq.~(\ref{eq:dphi})] both as (a) function of $T$ and (b) function of $\epsilon$. Particularly, in (b), both main figure and inset show that the value of $\mathrm{d}\epsilon \ll \epsilon$ for all $\epsilon$.}
\label{fig:dphi}
\end{figure}

\begin{figure*}
  \centering
  \subfigure[]{\includegraphics[scale=0.38]{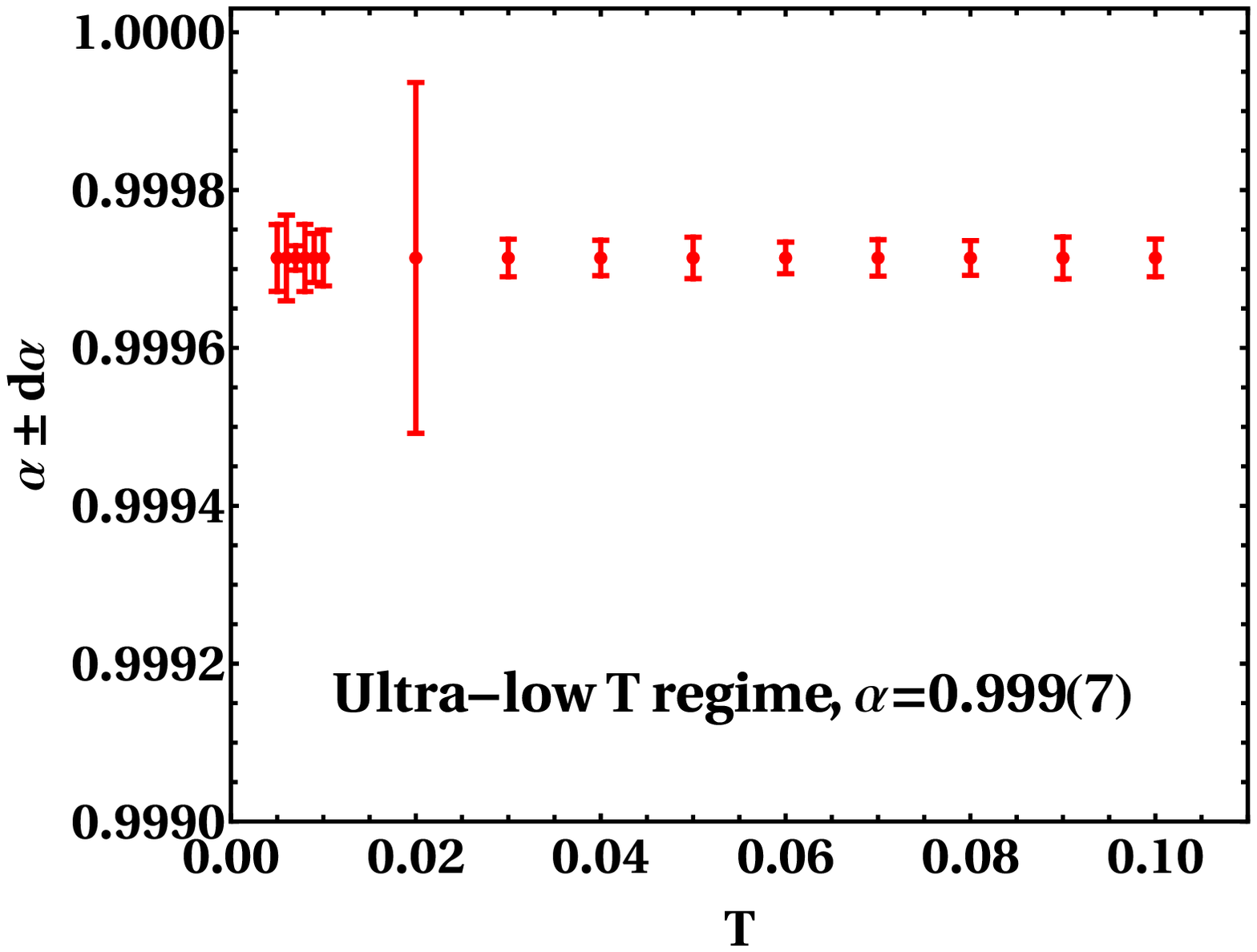}}\hfill
  \subfigure[]{\includegraphics[scale=0.38]{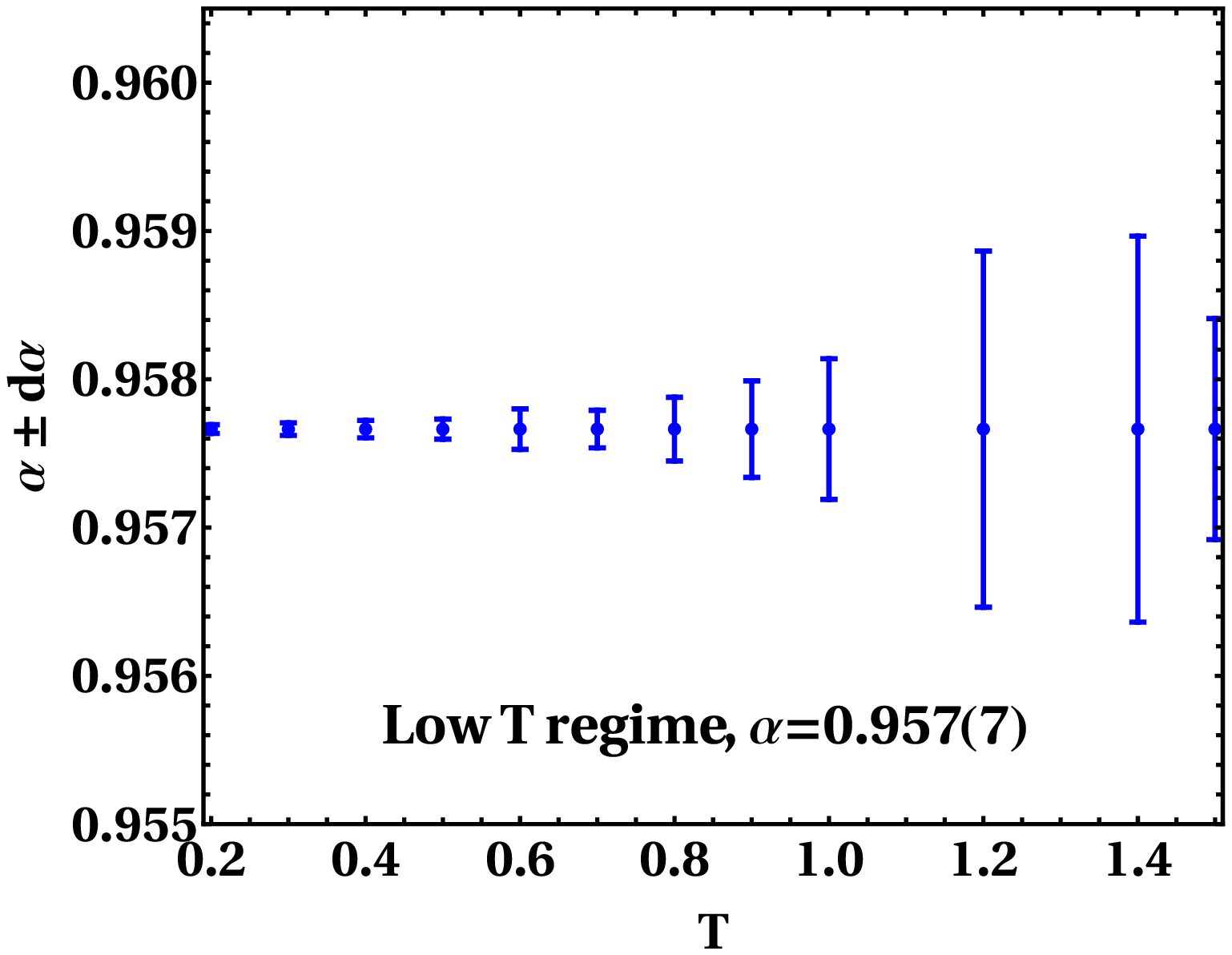}}\hfill
  \subfigure[]{\includegraphics[scale=0.38]{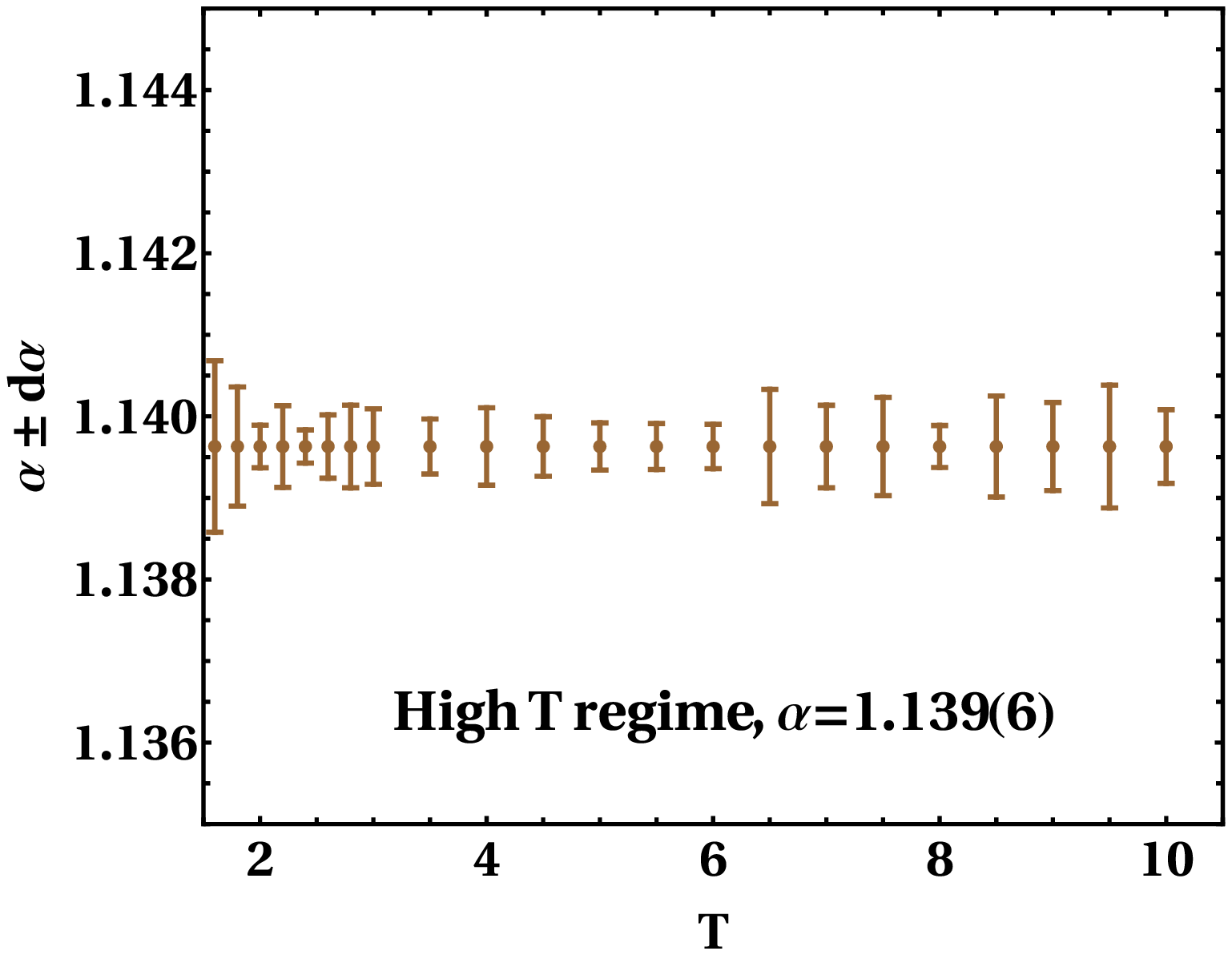}}\\
  \caption{The values of $\alpha$ in different dynamical regimes are plotted along with the corresponding error bars $\mathrm{d}\alpha$ 
  [Eq.~(\ref{eq:dalpha})]. In all three regimes, namely (a) ultra-low temperature regime, (b) low temperature regime and (c) high temperature regime, we observe that the corresponding error bars are significantly smaller than the $\alpha$ values themselves. Therefore, $\alpha$ can  prominently demarcate the three dynamical regimes.}
\label{fig:dalpha}
\end{figure*}

\item If we could repeat this numerical experiment for a large number of samples, we would get a sample average $\bar{\epsilon}_j$ for each of them. This way of doing enough repeats essentially implies averaging over the exact distribution of the observable. As explained in Ref.~\onlinecite{Young_2014}, it turns out that the sample mean $\bar{\epsilon}_j$ is the best estimate for the actual mean which is obtained in principle by using the exact probability distribution. However, this is expected to be accompanied with an error bar (standard deviation) $\bar{\sigma}_j$. The actual standard deviation $\bar{\sigma}_j$ is connected to the sample standard deviation $\bar{s}_j$ as \cite{Young_2014}
\be
\bar{\sigma}_j=\frac{\bar{s}_j}{\sqrt{n-1}}.
\label{eq:sigma}
\ee
So, we numerically compute $\bar{\sigma}_j$ for several  temperatures in the range of interest i.e. $0.005\leq T\leq 10 $. To get some estimates of the corresponding values, we present $\bar{\sigma}_j$ in Fig.~\ref{fig:sigmaj} at the two extreme temperatures $T=0.005$ (ultra-low temperature regime) and $T=10.0$ (high temperature regime). 

\item We have previously observed [Fig.~\ref{fig:vir_can}(f), Fig.~\ref{fig:sigmaj}(b)] spatial fluctuations in the spatial profiles of the average and standard deviation of energy at several temperatures. Then, it would be natural to perform further spatial averages over $N$ sites as follows 
\bea
\epsilon&=&\frac{1}{N} \sum_{j=1}^{N}\bar{\epsilon}_j\cr
\sigma&=&\frac{1}{N}\sum_{j=1}^{N}\bar{\sigma}_j.
\label{eq:sigmabar}
\eea
 Both $\epsilon$ and $\sigma$ are now independent of space (global) and they are functions of temperature. To get an estimate of $\sigma$ as a function of temperature, we present the corresponding plot in Fig.~\ref{fig:sigmaT}. 

\item In this connection, we define $\zeta=\sigma/\epsilon$ to measure the relative error with respect to the average. In Fig.~\ref{fig:nuT}, we observe  that for any temperature $\zeta<10^{-2}$. \\

\item It is important to note that, after the time average with respect to the sample of $n$ data points [Eq.~(\ref{eq:s}), Eq.~(\ref{eq:sigma})], the data set actually takes the form 
\be
\hspace*{0.5 cm}\left\lbrace y_1,\dots, y_N \right\rbrace=\left\lbrace \bar{\epsilon}_1 \pm \bar{\sigma}_1, \bar{\epsilon}_2 \pm \bar{\sigma}_2 \dots \bar{\epsilon}_N \pm \bar{\sigma}_N \right\rbrace.
\label{eq:yj}
\ee
To be precise, we have data points along with some error bars in Eq.~(\ref{eq:yj}). The average $\langle y \rangle$ is given by
\be
\hspace*{0.2 cm}\langle y \rangle=\frac{1}{N}\sum_{j=1}^{N}=\frac{1}{N}\sum_{j=1}^{N} (\epsilon_j\pm\sigma_j)=\epsilon\pm\sigma,
\ee
where we have used Eq.~(\ref{eq:sigmabar}). Consequently, using $\zeta=\sigma/\epsilon$, we have the following bound, 
\be
\epsilon(1-\zeta)\leq \langle y\rangle \leq \epsilon(1+\zeta).
\label{eq:avy}
\ee
Clearly, lesser the value of $\zeta$, better the convergence of $\epsilon$ to $\langle y\rangle$. This is indeed the case here since $\zeta<10^{-2} \ll 1$ as shown in Fig.~\ref{fig:nuT}.

\item Now, we look at the standard deviation $\bar{s}_y$ of the sample data set $y_j$-s in Eq.~(\ref{eq:yj}). One can show that
\be
\hspace*{0.5 cm}\bar{s}_y^2=\bar{s}_{\bar{\epsilon}}^2+\bar{s}_{\bar{\sigma}}^2\pm \frac{2}{N}\left[\sum_{j=1}^{N}\bar{\epsilon}_j\bar{\sigma}_j -\frac{1}{N}\sum_{j,k}\bar{\epsilon}_j\bar{\sigma}_k\right].
\ee

\item As already explained in Eq.~(\ref{eq:sigma}), the actual standard deviation $\bar{\sigma}_y$ here is connected to the sample standard deviation $\bar{s}_y$ as
\be
\bar{\sigma}_y=\frac{\bar{s}_y}{\sqrt{N-1}}.
\ee
Clearly, this $\bar{\sigma}_y$ is our required standard deviation $\mathrm{d}\epsilon$ i.e. 
\be
\mathrm{d}\epsilon=\bar{\sigma}_y.
\label{eq:dphi}
\ee
We present the estimates of $\mathrm{d}\epsilon$ in Fig.~\ref{fig:dphi} and observe that $\mathrm{d}\epsilon$ is sufficiently smaller than 
$\epsilon$ (for all temperatures).

\item Finally, we replace $\mathrm{d}\epsilon$ from Eq.~(\ref{eq:dphi}) in Eq.~(\ref{eq:dalpha}) to get the required error bar $\mathrm{d}\alpha$ in all the temperature regimes. 

\end{enumerate}
Having presented the detailed procedure above (points $1$ to $10$), we now plot the $\alpha$ values with the corresponding error bars $\mathrm{d}\alpha$ in different temperature regimes (Fig.~\ref{fig:dalpha}). The figure shows that indeed the error bars are sufficiently small and therefore the exponent $\alpha$ undoubtedly can serve as a prominent demarcator of the three different dynamical regimes.


\begin{thebibliography}{99}

\bibitem{Ablowitz_2004} Ablowitz M J, Prinari B and Trubatch A D, 2004, {\it Discrete and Continuous Nonlinear Schrödinger Systems}
(Cambridge: Cambridge University Press).

\bibitem{Kevrekidis_2001} P. G. Kevrekidis, K. Ø. Rasmussen and A. R. Bishop, Int. J. Mod. Phys. B {\bf 15}, 2833 (2001).

\bibitem{Hennig_1999}D. Hennig and G. P. Tsironis, Physics Reports {\bf 307}(5-6), 333 (1999).

\bibitem{Christodoulides_1988} D. N. Christodoulides and R. J. Joseph, Opt. Lett. {\bf 13}, 794 (1988).

\bibitem{Eisenberg_1998} H. Eisenberg, Y. Silberberg, R. Morandotti, A. Boyd and J. Aitchison, Phys. Rev. Lett. {\bf 81}, 3383 (1998).

\bibitem{Eisenberg_2000} H. Eisenberg, Y. Silberberg, R. Morandotti, and J. Aitchison, Phys. Rev. Lett. {\bf 85}, 1863 (2000).

\bibitem{Morandotti_1999} R. Morandotti, U. Peschel, J. Aitchison , H. Eisenberg, and Y. Silberberg, Phys. Rev. Lett. {\bf 83}, 2726 (1999).

\bibitem{Davydov_1973} A. S. Davydov, J. Theor. Biol. {\bf 38}, 559 (1973).

\bibitem{Davydov_1981} A.S. Davydov, Physica D {\bf 3}, 1 (1981).

\bibitem{Sievers_1988} A. J. Sievers and S. Takeno, Phys. Rev. Lett. {\bf 61}, 970 (1988).

\bibitem{Su_1979} W. P. Su, J. R. Schieffer, and A.J. Heeger, Physics. Rev. Lett. {\bf 42}, 698 (1979).

\bibitem{Trombettoni_2001} A. Trombettoni and A. Smerzi, Phys. Rev. Lett {\bf 86}, 2353 (2001). 

\bibitem{Rasmussen_2000} Rasmussen K Ø, Cretegny T, Kevrekidis P G and Grønbech-Jensen N, Statistical mechanics of a discrete nonlinear system, Phys. Rev. Lett. {\bf 84}, 3740 (2000).

\bibitem{Iubini_2013_NJP} S. Iubini, R. Franzosi, R Livi, G-L. Oppo and A. Politi, New J. Phys. {\bf 15}, 023032 (2013).

\bibitem{Iubini_2014} S. Iubini, A. Politi and P. Politi, J. Stat. Phys {\bf 154}, 1057 (2014).

\bibitem{Iubini_2012} Iubini S, Lepri S, and Politi A, Phys. Rev. E  {\bf 86}, 011108 (2012).

\bibitem{Iubini_2013} Iubini S, Lepri S, Livi R and Politi A, J. Stat. Mech.  {\bf 2015}, P08017 (2013).

\bibitem{Mendl_2015} C. Mendl and H. Spohn, J. Stat. Mech.  {\bf 2015}, P08028 (2015).

\bibitem{Kulkarni_2013} M. Kulkarni and A. Lamacraft, Phys. Rev. A {\bf 88}, 021603 (2013).

\bibitem{Prahofer_2004} M. Pr\"{a}hofer and H. Spohn, J. Stat. Phys. {\bf 115}, 255 (2004).

\bibitem{Kulkarni_2015} M. Kulkarni, D. A. Huse and H. Spohn, Phys. Rev. A {\bf 92}, 043612 (2015).

\bibitem{Gradenigo_2019} G. Gradenigo, S. Iubini, R. Livi and S. N. Mjumdar, arXiv:1910.07461 (2019).

\bibitem{Touchette_2015} H. Touchette, J. Stat. Phys {\bf 159}, 987 (2015).

\bibitem{Prigogine_1991} I. Prigogine, T. Y. Petroski, H. H. Hasegawa and S. Tasaki, Chaos  Solitons Fractals {\bf 1}, 3 (1991).

\bibitem{Masoliver_2011} J. Masoliver and A. Ros, Eur. J. Phys. {\bf 32}, 431 (2011).

\bibitem{Das_2018} A. Das, S. Chakrabarty, A. Dhar, A. Kundu, D. A. Huse, R. Moessner, S. S. Ray, and S. Bhattacharjee, Phys. Rev. Lett. {\bf 121}, 024101 (2018).

\bibitem{Bilitewski_2018} T. Bilitewski, S. Bhattacharjee, and R. Moessner, Phys. Rev. Lett. {\bf 121}, 250602 (2018).

\bibitem{Kumar_2019} D. Kumar, S. Bhattacharjee, and S. S. Ray, arXiv:1906.00016 (2019).

\bibitem{Chatterjee_2020} A K. Chatterjee, A. Kundu and M. Kulkarni, Phys. Rev. E {\bf 102}, 052103 (2020).

\bibitem{Ruidas_2020} S. Ruidas and S. Banerjee, arXiv:2007.12708 (2020).

\bibitem{Bhanu_2020} Bhanu K. S., D. A. Huse and M. Kulkarni, arXiv:2011.09320 (2020).

\bibitem{Bilitewski_2020} T. Bilitewski, S. Bhattacharjee and R. Moessner, arXiv:2011.04700 (2020).

\bibitem{Hasegawa_1973} A. Hasegawa and F. Tappert, Appl. Phys. Lett. {\bf 23}, 171 (1973).

\bibitem{Gardiner_2004} Gardiner C. W., 2004, {\it Handbook of Stochastic Methods}, Springer-Verlag Berlin Heidelberg, Germany.

\bibitem{Pathria_1986} Pathria R., 1986, {\it Statistical mechanics}, International Series in Natural Philosophy.

\bibitem{Howard_2005} J. E. Howard, Celestial Mechanics and Dynamical Astronomy {\bf 92}, 219 (2005).

\bibitem{Jin_1996} A. J. Jin, M. Veum, T. Stoebe, C. F. Chou, J. T. Ho, S. W. Hui, V. Surendranath and C. C. Huang, Phys. Rev. E {\bf 53}, 3639 (1996).

\bibitem{Chou_1997} C. F. Chou and J. T. Ho, Phys. Rev. E {\bf 56}, 592 (1997).

\bibitem{Chou_1998} C. F. Chou, A. J. Jin, S. W. Hui, C. C. Huang and J. T. Ho, Science {\bf 280}, 1424 (1998).

\bibitem{Touchette_2020} V. Drouin-Touchette, P. P. Orth, P. Coleman, P. Chandra and T. C. Lubensky, arXiv:2103.01878 (2020).

\bibitem{Das_2020} A. Das, K. Damle, A. Dhar, D. A. Huse, M. Kulkarni, C. B. Mendl and H. Spohn, J. Stat. Phys. {\bf 180}, 238 (2020).

\bibitem{Das_2015} S. G. Das and A. Dhar, arXiv:1411.5247 (2015).

\bibitem{Spohn_2014} H. Spohn, arXiv:1411.3907 (2014).

\bibitem{Kumar_2020} M. Kumar, A. Kundu, M. Kulkarni, D. A. Huse and A. Dhar, Phys. Rev. E {\bf 102}, 022130 (2020).

\bibitem{Sarma_2014} A. Sarma, M. Miri, Z. H. Musslimani and D. N. Christodoulides, Phys. Rev. E {\bf 89}, 052918 (2014). 

\bibitem{Ablowitz_2014} M. J. Ablowitz and Z. H. Musslimani, Phys. Rev. E {\bf 90}, 032912 (2014).

\bibitem{Ablowitz_2016} M. J. Ablowitz and Z. H. Musslimani, Studies in Applied Mathematics {\bf 139}, 7-59 (2016).

\bibitem{Mithun_2021} T. Mithun, A. Maluckov, B. M. Manda, Ch. Skokos, A. Bishop, A. Saxena, A. Khare and P. G. Kevrekidis, Phys. Rev. E {\bf 103}, 032211 (2021).

\bibitem{Ishimori_1982} Y. Ishimori, Journal of the Physical Society of Japan {\bf 51}, 3417 (1982).

\bibitem{Roberts_1988} J. A. G. Roberts and C. J. Thompson, J. Phys. A: Math. Gen. {\bf 21}, 1769 (1988).

\bibitem{Nowak_2015} U. Nowak, 2007, {\it Handbook of magnetism and advanced magnetic materials}, (Chichester: Wiley). 

\bibitem{Mannella_2002} R. Mannella, International Journal of Modern Physics C {\bf 13}(09), 1177 (2002).

\bibitem{Young_2014} P. Young, arXiv:1203.3781 (2014).

\end{thebibliography}
\end{document}